\documentclass[12pt,a4paper]{article}
\usepackage[dvips]{epsfig}
\usepackage{amssymb}
\begin{document}

\date{}
\title{
{\vspace{-2em} \normalsize 
{\hspace{1.5cm}November, 2001} \hfill \parbox[t]{50mm}{DESY 01-141\\MS-TP-01-15\\ROM2F/2001/36}}\\[10mm]
{\Large\bf  The supersymmetric 
Ward identities \\ on the lattice}                  \\[2mm]}
\author{\normalsize DESY-M\"unster-Roma Collaboration                         \\[3mm]
\normalsize
F.~Farchioni
\footnote{Address after October 1st: 
Institut f\"ur Theoretische Physik, Universit\"at M\"unster,
Wilhelm-Klemm-Str. 9, D-48149 M\"unster, Germany.}, 
C. Gebert, R.~Kirchner\footnote{Address after October 1st:
Universidad Aut\`onoma de Madrid, Cantoblanco, Madrid 28049, Spain.}, 
I.~Montvay,             \\
\normalsize
Deutsches Elektronen-Synchrotron DESY, D-22603 Hamburg, Germany \\[2mm]
\normalsize
A.~Feo\footnote{Work supported by the Deutsche Forschungsgemeinschaft (DFG).
Address after November 1st: School of Mathematics,
Trinity College, Dublin 2, Ireland.}, 
G.~M\"unster,              \\
\normalsize
Institut f\"ur Theoretische Physik,                      
Universit\"at M\"unster, Wilhelm-Klemm-Str.\,9, \\
\normalsize              
D-48149 M\"unster, Germany                                    \\[2mm]
\normalsize
T.~Galla,  \\
\normalsize
Department of Physics, University of Oxford, 
1 Keble Road, \\
\normalsize
Oxford OX1 3NP, UK \\[3mm]
\normalsize
A. Vladikas,              \\
\normalsize
INFN, Sezione di Roma 2, c/o Dipartimento di Fisica, \\
\normalsize
Univ. di Roma ``Tor Vergata'', 
Via della Ricerca Scientifica 1, 
I-00133 Rome, Italy.}


\newcommand{\LSB}{\raisebox{-0.3ex}{\mbox{\LARGE$\left[\right.$}}}
\newcommand{\RSB}{\raisebox{-0.3ex}{\mbox{\LARGE$\left.\right]$}}}
\newcommand{\blambda}{{\overline{\lambda}}}
\newcommand{\be}{\begin{equation}}
\newcommand{\ee}{\end{equation}}
\newcommand{\tr}{{\rm Tr}}
\newcommand{\btheta}{{\overline{\theta}}}
\newcommand{\hmu}{{\hat{\mu}}}
\newcommand{\hnu}{{\hat{\nu}}}
\newcommand{\vev}[1]{\left\langle #1 \right\rangle}
\newcommand{\one}{1\!\!1}
\newcommand{\ccc}{\textsf{C}}
\newcommand{\ppp}{\textsf{P}}
\newcommand{\ttt}{\textsf{T}}
\newcommand{\spp}{\textsf{\scriptsize P}}
\newcommand{\stt}{\textsf{\scriptsize T}}
\newcommand{\scc}{\textsf{\scriptsize C}}
\newcommand{\m}{\hphantom{$-$}}
\newcommand{\z}{\hphantom{$0$}}

\begin{titlepage}  

\maketitle
\thispagestyle{empty} 

\newpage

\thispagestyle{empty} 
\begin{abstract} \normalsize

Supersymmetric (SUSY) Ward identities are considered
for the N=1 SU(2) SUSY Yang Mills theory discretized 
on the lattice with Wilson fermions (gluinos).
They are used in order to compute non-perturbatively
a subtracted gluino mass and the mixing coefficient of the SUSY
current. 
The computations were performed at gauge coupling $\beta=2.3$
and hopping parameter $\kappa\!=\!0.1925$, 0.194, 0.1955
using the two-step multi-bosonic dynamical-fermion
algorithm.
Our results are consistent with a scenario where the Ward 
identities are satisfied up to $O(a)$ effects.
The vanishing of the gluino mass occurs at a value of the
hopping parameter which is not fully consistent with the estimate based on 
the chiral phase transition.
This suggests that, although SUSY restoration appears to occur
close to the continuum limit of the lattice theory, the results are
still affected by significant systematic effects.

\end{abstract}

\end{titlepage}  

\newpage

\section{Introduction}

A better understanding of non-perturbative phenomena 
in supersymmetric (SUSY) gauge theories could be gained in the framework of
the lattice regularization. 
An immediate difficulty arises, however, because the lattice regularized
theory is not supersymmetric as the Poincar\'e invariance, a
sector of the superalgebra, is lost.
This is evident if one considers the super-algebra in the canonical formalism 
(the notation is with Weyl spinors)
\be
\{Q_\alpha,\bar{Q}_\beta\}\:=\:2\,\sigma_{\alpha\beta}^{\mu}\,P_\mu\ .
\nonumber\ee
This relation cannot be fulfilled in a
discrete space-time manifold, where momenta are not
generators of infinitesimal space-time translations.
More specific difficulties arise in the fermionic sector where
spurious states may violate the balance between bosonic 
and fermionic degrees of freedom. In the standard approach 
to lattice gauge theories with Wilson fermions 
(for an approach using domain-wall fermions
see \cite{Col}) the suppression of spurious states in
the continuum limit is obtained by adding  to the action an 
`irrelevant' term (Wilson term) 
which explicitly breaks SUSY.
 
Some time ago Curci and Veneziano \cite{CuVe} proposed 
that in spite of this substantial SUSY breaking 
the Wilson discretization may be safely applied to 
SUSY gauge theories: the symmetry is recovered 
in the continuum limit by properly tuning the bare parameters of the action.
They considered the simple example of the 
N=1 supersymmetric Yang Mills theory (SYM). 
This is the supersymmetrized version of quantum gluodynamics 
where the $N_c^2-1$ gluons are accompanied by 
an equal number of fermionic partners (gluinos) in the same (adjoint) 
representation of the color group.
The Wilson action for the N=1 SYM breaks SUSY by 
the Wilson term for the gluino action and by a gluino Majorana mass
term. 
The lattice Ward identities (WIs), considered in \cite{BoetAl}
for the case of the chiral symmetry in QCD (see also \cite{KaSm,Testa}), 
provide a general theoretical framework for properly
dealing with the problem of the restoration 
of the symmetry in the continuum limit.
As a consequence of the SUSY breaking, 
similarly to QCD, the gluino mass is shifted: a tuning 
procedure on the bare mass is required in order to  recover
massless gluino and SUSY in the continuum limit. 

The problem of the Monte Carlo simulation of the N=1 SU(2) SYM 
with Wilson fermions was considered by this collaboration in the past
\cite{Mont2,KoetAl,KietAl,KietAl2,CaetAl} (for a study in the quenched approximation 
see \cite{DoetAl}).
The theory with dynamical gluino was simulated  by the two-step 
multi-bosonic (TSMB) algorithm defined in \cite{Mont3,Mont4}. 
The formulation of the algorithm, flexible with respect 
to the parameter $N_f$, 
allows to treat the gluino which is a Majorana fermion ($N_f\!=\!1/2$). 
An extensive analysis of the low-energy aspects of the SU(2) SYM
was performed.

An intricate point in the N=1 SYM is the tuning of the theory to the 
massless gluino limit. The straightforward spectroscopic method
familiar in QCD cannot be applied in the case of the  N=1 SYM 
since the theory possesses only 
an anomalous axial chiral symmetry and no Goldstone boson is available.
The problem was considered in~\cite{KietAl2}
by studying the behavior of the finite-volume gluino condensate 
as a function of the hopping parameter: the massless gluino 
is expected to correspond to two degenerate vacua 
with symmetric probability distribution of the gluino condensate.
SUSY restoration can be also verified by direct
inspection of the low-energy mass spectrum~\cite{CaetAl}: 
this is expected to reproduce the SUSY multiplets 
predicted by the low-energy effective 
Lagrangians \cite{VeYa,FaGaSch}. An accurate analysis of the 
spectrum is however a non-trivial task from the computational point of view
and an independent method for checking SUSY is welcome.
A possibility~\cite{DoetAl} is to use the SUSY WIs
to determine the gluino mass
in the same way the chiral WIs are used in QCD
to determine the quark mass. 
A great simplification consists in considering the on-shell regime. 
In addition the WI approach improves the
insight in the renormalization of the lattice SUSY current.
The properly renormalized current defines the supercharge
and satisfies the appropriate superalgebra.
The renormalization of the lattice SUSY current is more complicated 
compared to the chiral case since SUSY is more severely broken
on the lattice.
Explicit one-loop calculations in lattice perturbation theory 
may lead to a better understanding of this problem~\cite{Tani,pert}.

In this work we concentrate on the SUSY WI approach.
As we shall see, two unknown parameters appear in the on-shell SUSY WIs.
These are ratios of three coefficients entering the WIs: $Z_S$ and $Z_T$ 
multiplying the divergences of the SUSY current $\partial_\mu S_\mu(x)$
and of the mixing current $\partial_\mu T_\mu(x)$, and 
the subtracted gluino mass $m_S$.
For $m_S=0$, SUSY is expected to be restored in the continuum limit.
As a result of the study we obtain a non-perturbative determination of
the dimensionless ratios $am_S Z_S^{-1}$ and $Z_T Z_S^{-1}$. We consider the gauge coupling 
$\beta\!=\!2.3$ on a $12^3\times24$ lattice
and three values of the hopping parameter 
$\kappa\!=\!0.1925$, 0.194, 0.1955, corresponding to 
decreasing gluino mass.
Preliminary results were presented in~\cite{proc1,Kir00,proc2}.

The paper is organized as follows.
In Sec.~2 the Curci-Veneziano approach is introduced in
the case of the N=1 SYM. 
The lattice action is defined and its symmetries are pointed out
(see also Appendix~A).
The latter play a special r\^ole in the analysis of the SUSY WIs.
In Sec.~3 the formalism of the lattice WIs is reviewed.
The fundamental issue in this context is the renormalization
of the `irrelevant' operator entering the WIs because of the explicit breaking
of SUSY in the lattice model. 
A detailed account is given in Appendix~B 
with an analysis based on the discrete hypercubic group.
The result is that the SUSY WIs assume a specific lattice form.
We discuss in this paper only the simplified case 
of the on-shell regime. 
Suitable gluino-glue insertion operators are discussed in Sec.~4. 
In Appendix~C the r\^ole of the symmetries  in this context is clarified.
In Sec.~5 we give an account of the present setup of the 
TSMB algorithm.
The parameters have been tuned in order to get good performance
for light fermionic degrees of freedom.
We also measure some quantities (smallest eigenvalue, 
sign of the Pfaffian, the scale $r_0$) which characterize the 
set of configurations under study.
Sec.~6 is devoted to the numerical analysis of the SUSY WIs
with an outline of the method and the presentation of the numerical results.
Conclusions are finally drawn in Sec.~7.

\section{The N=1 SYM on the lattice} 

We adopt the formulation of Ref.~\cite{CuVe} for the lattice 
discretization of the N=1 SYM with $N_c$ colors.
The pure gauge action $S_g$ is the standard plaquette one
\be  \label{gauge}
S_g  = \frac{\beta}{2} \sum_x\sum_{\mu\neq\nu}                            
\left( 1 - \frac{1}{N_c} {\rm Re\,Tr\,} U_{\mu\nu}(x) \right) \ ,   
\ee
where the plaquette is defined as
\be  \label{plaqact}
U_{\mu\nu}(x)\: =\: U_{\nu}^\dagger(x)\, U_{\mu}^\dagger(x+\hnu)\,
U_{\nu}(x+\hmu)\, U_{\mu}(x)\ ,
\ee
and the bare gauge coupling is given by $\beta \equiv 2N_c/g_0^2$.
For Wilson fermions the action (with Wilson parameter set to $r\!=\!1$) reads
\begin{eqnarray}  \label{fermi1}
S_f&=&
\sum_x\, a^4\,\tr \LSB\,
 \frac{1}{2a}\sum_\mu\left(\blambda(x)\, (\gamma_\mu-1)\,
 U^\dagger_\mu(x)\, \lambda(x+\hmu)\, U_\mu(x)\,\right.
\nonumber\\&& -\: \left.
\blambda(x+\hmu)\, (\gamma_\mu+1)\,
 U_\mu(x)\, \lambda(x)\, U^\dagger_\mu(x)\right)\,
\,+\,(m_0+\frac{4}{a})\, \blambda(x)\lambda(x)\,
\RSB\ ,
\end{eqnarray}
with $a$ the lattice spacing and $m_0$ the gluino bare mass.
The gluino field $\lambda(x)$ is a Majorana spinor transforming according 
to the adjoint representation of the gauge group.
The symbol `Tr' denotes the trace over the color indices. 
In this work we consider $N_c\!=\!2$, for which the adjoint
gluino field is expressed in terms of Pauli matrices  $\sigma_r$
\be 
\lambda=\sum_{r=1}^3\frac{1}{2}\sigma_r\lambda^r\ .
\ee 
The following relation (Majorana condition) holds for an Euclidean Majorana field:
\be\label{ettore}
\lambda(x)=\lambda^\scc(x)\equiv C\blambda^{T}(x)\ ,
\ee
where $C=\gamma_0\gamma_2$ is the spinorial matrix associated 
with the charge-conjugation symmetry~\ccc.
Boundary conditions are taken to be periodic except
for fermionic fields in the time direction, which are anti-periodic.

In Monte Carlo simulations a different parametrization  
is used. The hopping parameter $\kappa$ is defined as
\be   \label{hopping}
\kappa = \frac{1}{2(4+m_0a)}\ 
\ee
and the fermionic action is expressed in terms of the fermion  matrix $Q$
\be \label{MCaction}
S_f\:=\:\frac{a^3}{2\kappa}
\sum_{xr,ys}\blambda^s (y) Q_{ys,xr}\lambda^r (x)
\ee
(Dirac indices are implicit).
The fermion matrix is given by
\be  
\label{faction}
Q_{ys,xr} \equiv Q_{ys,xr}[U] \equiv
\delta_{yx}\delta_{sr} - \kappa \sum_{\mu=1}^4 \left[
\delta_{y,x+\hat{\mu}}(1+\gamma_\mu) V_{sr,x\mu} + 
\delta_{y+\hat{\mu},x}(1-\gamma_\mu) V^T_{sr,y\mu} \right] 
\ee
with the adjoint link $V_{rs,x\mu}(x)$ defined as
\be
\label{alink}
V_{rs,x\mu} \equiv V_{rs,x\mu}[U] \equiv
\frac{1}{2}\, {\rm Tr}(U_{x\mu}^\dagger \sigma_r U_{x\mu} \sigma_s)
= V_{rs,x\mu}^* =V_{rs,x\mu}^{-1T} \ .
\ee

The action~(\ref{gauge})-(\ref{fermi1}) is invariant under
the discrete symmetries \ppp\ (parity), \ttt\ 
(time-reversal) and \ccc\ (charge-conjugation).
For the case under consideration the latter symmetry
implies the following relation for the fermion matrix
\be
Q^T_{xr,ys}[U]=CQ_{ys,xr}[U]C^{-1}
\ee
(transposition is intended on the suppressed Dirac indices).
The discrete symmetries of the lattice action play an important 
r\^ole in the subsequent analysis of the
WIs; their explicit definition is given in Appendix~A.

\section{Lattice SUSY Ward identities} 
\label{sec:lwi}

SUSY is explicitly broken in the action~(\ref{gauge})-(\ref{fermi1}) 
by the gluino mass term, by the Wilson term and by the lattice discretization.
Using lattice SUSY WIs a soft-breaking subtracted gluino mass $m_S$
can be defined. 
The expectation is that the vanishing of $m_S$, 
for asymptotically small lattice spacings, 
ensures the restoration of SUSY up to discretization effects.
In this Section we discuss these issues, which were first introduced
in \cite{CuVe}, and have also been considered in \cite{DoetAl} 
and \cite{Tani}. 

Lattice SUSY transformations complying with gauge invariance,
\ppp, \ttt\ and the Majorana nature of the gluino field 
are \cite{Gal,Tani}\footnote{Our definition of
the link variable $U_\mu(x)$ differs from that 
of~\cite{Tani} (see our definition of the plaquette~(\ref{plaqact})); 
the two definitions are related by Hermitian conjugation.}:
\begin{eqnarray}
\nonumber
\delta U_\mu(x) &=&
 -\frac{ig_0a}{2} (\btheta(x) \gamma_\mu 
U_\mu(x) \lambda(x) + \btheta(x+\hmu)\gamma_\mu\lambda(x+\hmu) U_\mu(x))\ ,
\nonumber\\\nonumber
\delta U^{\dagger}_\mu(x) &=&
\frac{ig_0a}{2} (\btheta(x) \gamma_\mu 
\lambda(x) U^{\dagger}_\mu(x) + \btheta(x+\hmu) \gamma_\mu U^{\dagger}_\mu(x) \lambda(x+\hmu))\ ,
\\\nonumber
\delta \lambda(x) &=&
\frac{1}{2} P^{(cl)}_{\mu\nu}(x) \sigma_{\mu\nu} \theta(x)\ ,
\\\label{susytr}
\delta  \blambda(x) &=&
-\frac{1}{2} \btheta(x) \sigma_{\mu\nu} P^{(cl)}_{\mu\nu}(x)\ ,
\end{eqnarray}
where $\theta (x)$, $\btheta (x)$ are  infinitesimal Majorana fermionic parameters.
The lattice field tensor $P^{(cl)}_{\mu\nu} (x)$ is 
clover-symmetrized so as to comply with \ppp\ and \ttt:  
\be\label{clover}
P^{(cl)}_{\mu\nu}(x) = \frac{1}{4a} \sum_{i=1}^4 \frac{1}{2ig_0a}
\left(U^{(i)}_{\mu\nu}(x)-U^{(i)\dagger}_{\mu\nu}(x)\right), 
\ee
where
\begin{eqnarray}
&&
U^{(1)}_{\mu\nu} (x) =
U^\dagger_{\nu}(x) U^\dagger_{\mu}(x+\hnu)
 U_{\nu}(x+\hmu) U_{\mu}(x)\equiv U_{\mu\nu}(x)\ , \nonumber
\\&&
U^{(2)}_{\mu\nu} (x) = U_{\mu}^\dagger(x) U_{\nu}(x+\hmu-\hnu)
 U_{\mu}(x-\hnu) U^\dagger_{\nu}(x-\hnu)\ , \nonumber
\\&&
U^{(3)}_{\mu\nu} (x) =
 U_{\nu}(x-\hnu) U_{\nu}(x-\hmu-\hnu)
 U^\dagger_{\mu}(x-\hmu-\hnu) U^\dagger_{\mu}(x-\hmu)\ , \nonumber
\\&&
U^{(4)}_{\mu\nu} (x) = U_{\mu}(x-\hmu) U^\dagger_{\nu}(x-\hmu)
 U^\dagger_{\mu}(x-\hmu+\hnu) U_{\nu}(x)\ .
\end{eqnarray}

For any operator $Q(y)$ the expectation value $\langle Q(y)\rangle$ 
is invariant if in the functional integral a change of variables 
according to the above SUSY transformations is performed. 
In the case of a gauge invariant operator $Q(y)$
this results in the following WI  
\begin{eqnarray} \label{ward1}
\sum_\mu\vev{\left(\nabla_\mu S^{(ps)}_\mu(x)\right) Q(y)}
&=&
m_0 \vev{\chi(x) Q(y)} + \vev{X^{(ps)}(x) Q(y)}
- \vev{\frac{\delta Q(y)}{\delta \btheta(x)}}\ .
\end{eqnarray}
The SUSY current $S^{(ps)}_\mu(x)$ is point-split ($ps$)~\cite{Tani}
\be\label{current1}
S^{(ps)}_\mu(x) =
-\frac{1}{2} \sum_{\rho\sigma} \sigma_{\rho\sigma} \gamma_\mu
 \tr \biggl(
 P^{(cl)}_{\rho\sigma}(x) U^\dagger_\mu(x)\lambda(x+\hmu)U_\mu(x)
+P^{(cl)}_{\rho\sigma}(x+\mu) U_\mu(x)\lambda(x)U^\dagger_\mu(x)
\biggr)\ ,
\ee
and the lattice derivative is the backward one
$\nabla^{b}_\mu f(x)=(f(x)-f(x-\hmu))/a$.
We recall that SUSY is broken by the presence of a non-zero bare mass
in the action, by the Wilson term and by the discretization.
The first type of SUSY breaking gives rise to the term of the WI~(\ref{ward1})
involving the operator $\chi(x)$:
\be\label{chi}
\chi(x) = \sum_{\rho\sigma} \sigma_{\rho\sigma}
 \tr \left(P^{(cl)}_{\rho\sigma}(x) \lambda(x)\right)\ .
\ee
The rest of the SUSY breaking results in the presence of 
the $X^{(ps)}(x)$ term. Its exact expression~\cite{Tani} 
is not needed in the following.
It suffices to know that in the naive continuum limit $X_S(x)\approx aO_{11/2}(x)$,
where $O_{11/2}(x)$ is a dimension-$11/2$ operator.

The last term in~(\ref{ward1}) is a contact term 
which vanishes when the distance $\vert x-y \vert$ is non-zero.
This corresponds to the {\em on-shell} situation. 
We shall restrict ourselves to this regime in the numerical analysis of the WIs
and contact terms will be consequently disregarded in the following discussions.

The definition of the SUSY current on the lattice is 
arbitrary up to terms which vanish in the continuum limit.
Another choice is the local ($loc$) current
\be\label{current2}
S^{(loc)}_\mu(x) =
-\sum_{\rho\sigma} \sigma_{\rho\sigma} \gamma_\mu
 \tr \biggl(
 P^{(cl)}_{\rho\sigma}(x) \lambda(x)
\biggr)\ .
\ee
This definition is preferable on the classical level \cite{Gal}
and is more convenient for analytic perturbative calculations. 
The local current $S^{(loc)}_\mu(x)$ satisfies a WI of the form~(\ref{ward1}), with 
a symmetric lattice derivative 
$\nabla^{s}_\mu f(x)=(f(x+\mu)-f(x-\hmu))/2a$ 
(required to preserve \ppp\ and \ttt) and a SUSY-breaking term
$X^{(loc)}= X^{(ps)} + O(a)$.

\subsection{Renormalization}
\label{sec:renorm}

The WI~(\ref{ward1}) is a  relation between bare 
correlation functions. The r\^ole of the symmetry-breaking operator $X^{(ps)}(x)$
(or $X^{(loc)}(x)$) is of particular interest, as it is related to current normalization
and gluino mass subtraction. Its treatment in the present case
follows closely that of the axial WIs in QCD~\cite{BoetAl,Testa}.

We consider the renormalization of the dimension-$11/2$ operator $O_{11/2}(x)$.
According to the usual prescriptions, this implies
mixing with operators of equal or lower dimensions $d$, which have
the same transformation properties under the
symmetries of the lattice action.
A discussion of the mixing pattern on the basis of the 
discrete hypercubic group is carried out in Appendix~B. The result is that
no Lorentz-breaking mixing arises at least in the on-shell regime.
The mixing pattern (involving operators with $7/2\!\leq\! d\leq 11/2$) 
in the on-shell case is given by
\begin{eqnarray}\nonumber
O^R_{11/2}(x)\:=\:Z_{11/2}
[O_{11/2}(x)+a^{-1}(Z_S-1)\,\nabla_\mu S_\mu(x)+a^{-1}Z_T\,\nabla_\mu T_\mu(x)+ 
a^{-2}Z_\chi\,\chi (x)] \\\label{renorm}
+\sum_j Z^{(j)}_{11/2}O^{(j)\:R}_{11/2}(x)\ .
\end{eqnarray}
Since it is not relevant to the present discussion we have left
unspecified the exact lattice form of the SUSY current 
(point-split or local) and derivative (backward or symmetric).
The same applies to the other dimension-$9/2$ 
operator appearing in Eq.~(\ref{renorm}), 
namely the divergence of the mixing current $T_\mu(x)$. It may be defined in analogy to 
$S_\mu(x)$ as point-split 
\be\label{T1}
T^{(ps)}_\mu(x) =
\sum_\nu \gamma_\nu 
 \tr \biggl(
 P^{(cl)}_{\mu\nu}(x) U^\dagger_\mu(x)\lambda(x+\hmu)U_\mu(x)
+P^{(cl)}_{\mu\nu}(x+\mu) U_\mu(x)\lambda(x)U^\dagger_\mu(x)
\biggr)\ 
\ee
or local
\be\label{T2}
T^{(loc)}_\mu(x) =
2\sum_{\mu\nu} \gamma_\nu
 \tr \biggl(
 P^{(cl)}_{\mu\nu}(x) \lambda(x)
\biggr)\ ,
\ee
with the lattice derivative chosen as in the case of the current $S_\mu(x)$.
From the above discussion obviously follows that different lattice currents $S_\mu(x)$ and 
$T_\mu(x)$ are associated with different values of $Z_S$ and $Z_T$.

The last term on the r.h.s. of Eq.~(\ref{renorm}) reflects the mixing
of the operator $O_{11/2}(x)$ with other bare operators
$O^{(j)}_{11/2}(x)$ of equal dimension. 
The reason Eq.~(\ref{renorm}) has been expressed in terms of the 
renormalized ones $O^{(j)\:R}_{11/2}(x)$ will become clear shortly.
The multiplicative renormalization  $Z_{11/2}$ and the 
mixing coefficients $Z^{(j)}_{11/2}$ are logarithmically divergent
in perturbation theory.
Solving Eq.~(\ref{renorm}) for $O_{11/2}(x)$ and substituting
it in WI~(\ref{ward1}) one gets
\be\label{renormward}
Z_S\vev{\left(\nabla_\mu S_\mu(x)\right) Q(y)}+
Z_T\vev{\left(\nabla_\mu T_\mu(x)\right) Q(y)}=
m_S \vev{\chi(x) Q(y)}\,+\,O(a)\ ,
\ee
where the subtracted mass $m_S$ is given by
\be
m_S = m_0-a^{-1}Z_{\chi}\ .
\ee
In deriving Eq.~(\ref{renormward}) we have relied on the vanishing
in the continuum limit of the correlation
\be\label{ordera}
a\,\vev{[Z^{-1}_{11/2}
O^R_{11/2}(x)-\sum_j 
Z^{(j)}_{11/2}O^{(j)\:R}_{11/2}(x)]\: Q(y)}\:=\:O(a)\ ,  
\ee
which is valid on-shell, $x\!\neq\! y$ 
(recall that $Z_{11/2}$, $Z^{(j)}_{11/2}$ are only logarithmically divergent).

By using general renormalization group arguments (see e.g. \cite{Testa}) one can show that
$Z_S$, $Z_T$ and $Z_\chi$, being power-subtraction coefficients,
do not depend on the renormalization scale
$\mu$ defining the renormalized operator~(\ref{renorm}).
Consequently dimensional considerations imply 
$Z_S = Z_S(g_0,m_0 a)$, $Z_T = Z_T(g_0,m_0 a)$,  
$Z_\chi = Z_\chi(g_0,m_0 a)$.
The requirement of a well defined chiral limit of the theory
implies in particular that the dependence of $Z_S$ and $Z_T$ on 
the gluino mass is vanishingly small in the continuum limit.
In simulations at fixed lattice spacing this dependence is 
treated as an $O(a)$ effect. 

In QCD the lattice chiral WI, a relation analogous to Eq.~(\ref{renormward}), leads to 
the definition of an axial current $\hat{A}_\mu(x)=Z_A A^{lat}_\mu(x)$ 
where $A^{lat}_\mu(x)$ is a generic discretization of the chiral current.
A rigorous argument~\cite{BoetAl,Testa} shows that the current $\hat{A}_\mu(x)$ 
coincides with the correctly normalized continuum chiral current. 
It satisfies the appropriate current algebra.
It is tempting to associate by analogy the quantity $\hat{S}_\mu(x)=Z_S S_\mu(x)+Z_T T_\mu(x)$ 
with the correctly normalized continuum SUSY current. 
An attempt to reproduce in this case the QCD argument fails. 
This is because the proof in QCD relies on two key properties
(cf. Sec.~3.1 of Ref.~\cite{Testa}):
\begin{enumerate}
\item The axial variation of the quark field is proportional to the field 
itself.
\item The gauge fixing term is invariant under axial transformations.
\end{enumerate}
The above statements, valid for the axial symmetry, do not apply to SUSY. 
Thus, to the best of our knowledge, SUSY WIs cannot be used in a
way analogous to the QCD chiral ones in order
to prove the non-renormalization theorem for the current $\hat{S}_\mu(x)$.
Explicit one-loop calculations in lattice perturbation theory 
may shed some light on this issue~\cite{Tani,pert}.
If the correctly normalized SUSY current coincides with $\hat{S}_\mu(x)$
(or is related to it by multiplicative renormalization),
it is conserved when $m_S$ vanishes.
This is the restoration of SUSY in the continuum limit.

\section{Insertion operators}
\label{sec:operators}

In this Section we turn our attention to the insertion operator
$Q(x)$ of the WI~(\ref{renormward}).
The operators  $\nabla_\mu S_\mu(x)$,
$\nabla_\mu T_\mu(x)$ and $\chi(x)$, which will be in the following
collectively denoted as `sink operators', transform according to
the bispinorial representation of the Poincar\'e group in the continuum.
In order to get a non-trivial WI, the insertion operator $Q(x)$
is required to contain (at least) one non-zero spin-$1/2$ component.
Thus, given a composite operator ${\cal O}$ which is a Majorana bispinor,
$Q(x)$ may be chosen to be of the form 
\be\label{obar}
Q(x)=\bar{\cal O}^T(x)\equiv C^{-1}{\cal O}(x) \,\,.
\ee
Clearly this operator must also be gauge invariant.

We consider the zero spatial momentum  WI
obtained by summation over the spatial coordinates
of Eq.~(\ref{renormward})
\begin{eqnarray} \nonumber
\sum_{\vec{x}}\vev{\left(\nabla_0 S_0(x)\right) \bar{\cal O}^T(y)}+
Z_T Z_S^{-1}\sum_{\vec{x}}\vev{\left(\nabla_0 T_0(x)\right) \bar{\cal O}^T(y)}=\\
\label{zeromomward}
m_S Z_S^{-1} \sum_{\vec{x}} \vev{\chi(x) \bar{\cal O}^T(y)}\ +O(a)\ .
\end{eqnarray}
Note that in the above equation the three correlation functions
are $4\times 4$ matrices in Dirac indices.
In numerical simulations these bare correlation functions
can be computed at fixed lattice bare lattice parameters $\beta=2N_c/g_0^2$ and $\kappa$.
Thus by choosing two elements of the $4 \times 4$ matrices, a system
of two equations can be solved for $m_S Z_S^{-1}$ and $Z_T Z_S^{-1}$.

One must clearly ensure that these two equations are non-trivial and
independent. To do this, let us consider the correlations 
containing the SUSY current (identical considerations apply for the other
two correlations). Written explicitly with its Dirac indices this reads
\be\label{corr_ex}
C^{(S,{\cal O})}_{\alpha\beta}(t)=a^{d_{\cal O}+9/2}\sum_{\vec{x}}\vev{(\nabla_0 S_0)_\alpha (x)\, 
\bar{\cal O}_\beta (y)}\ ,\quad\quad t=x_0-y_0\ .
\ee
We consider dimensionless correlations, since these are the quantities 
actually computed in simulations.
The above $4\times 4$ matrix can be expanded in the basis of the 
16 Dirac matrices $\Gamma$
\be
C^{(S,{\cal O})}_{\alpha\beta}(t)=\sum_\Gamma C^{(S,{\cal O})}_{\Gamma}(t)\,\Gamma_{\alpha\beta} \ .
\ee
Using discrete symmetries, see Appendix~C, we can show that the only
surviving contributions are (here Dirac indices are contracted)
\begin{eqnarray}\label{comp1}
C^{(S,{\cal O})}_{\one}(t)&\equiv&\sum_{\vec{x}}\vev{\left(\overline{\nabla_0 S}_0(x)
{\cal O}(y)\right)} \\ \label{comp2}
C^{(S,{\cal O})}_{\gamma_0}(t)&\equiv&\sum_{\vec{x}}\vev{\left(\overline{\nabla_0 S}_0(x)
\gamma_0{\cal O}(y)\right)}\ .
\end{eqnarray}
Due to the Majorana nature of the operators, the correlations
$C^{(S,{\cal O})}_{\one}(t)$ and $C^{(S,{\cal O})}_{\gamma_0}(t)$ are real.
In conclusion for a given insertion operator
we determine the dimensionless quantities $am_S Z_S^{-1}$ and $Z_T Z_S^{-1}$
by solving the system of two equations
\be\label{system}
\left\{
\begin{array}{l}
C_{\one}^{(S,{\cal O})}(t)\:\:+\:\:(Z_T Z_S^{-1})\:C_{\one}^{(T,{\cal O})}(t)\:\:=\:\: (am_S Z_S^{-1})\: 
C_{\one}^{(\chi,{\cal O})}(t)\\
C_{\gamma_0}^{(S,{\cal O})}(t)\:\:+\:\:(Z_T Z_S^{-1})\:C_{\gamma_0}^{(T,{\cal O})}(t)\:\:=\:\:(am_S Z_S^{-1})\:C_{\gamma_0}^{(\chi,{\cal O})}(t)\ .\\
\end{array}
\right.
\ee

We now turn our attention to the choice of suitable insertion
operators $Q(x)$.
Practical considerations suggest the use of the lowest-dimensional insertion 
operators with the suitable symmetry properties. In our case this means the
$d = 7/2$ gauge invariant bispinor
\be\label{insop}
\tr \left[P^{(cl)}_{\mu\nu}(x) \lambda(x)\right]
\ee
which is a tensor of 6 components in the Lorentz indices. 
Since the sink operators
have spin-$1/2$, we must project out of the above Lorentz tensor the 
spin-$1/2$ components. Examples are $S_0$, $T_0$, $\chi$ and
\be\label{phi1}
{\chi}^{(sp)}(x)=\sum_{i<j}\sigma_{ij}\tr\left[P^{(cl)}_{ij}\lambda(x)\right]
\ee
(only spatial plaquettes are taken into account).
Since the Lorentz tensor of Eq.~(\ref{insop}) has only two independent 
spin-$1/2$
components (see Appendix~C for a detailed discussion), 
not all of the above operators can be independent.  
Indeed, they are related by
\begin{eqnarray}\label{oprel1}
\chi(x)&=&\hphantom{2\gamma_0(}\gamma_0 T_0(x)-2{\chi}^{(sp)}(x) \\ \label{oprel2}
S_0(x)&=&2\gamma_0(\gamma_0T_0(x)-2{\chi}^{(sp)}(x))\ . 
\end{eqnarray} 
We see that two independent systems of 
equations~(\ref{system}) exist for two choices of dimension-$7/2$ insertion
operators $O$. This redundancy can in principle be used in order to check lattice artifacts.

\section{Simulation of the model with light gluinos}

We simulate the N=1 SU(2) SYM on a $12^3\times24$ lattice
at $\beta\!=\!2.3$. This value of $\beta$ corresponds to 
the lower end of the approximate scaling region in pure SU(2) 
lattice gauge theory. 
In the full theory virtual loops of gluinos 
contribute to the Callan-Symanzik $\beta$-function.
The consequence is that, for fixed $\beta$, the lattice spacing is decreased.

The scaling properties of the model with dynamical gluinos
were studied in detail in \cite{CaetAl}.
There values of $\kappa$ up to $\kappa\!=\!0.1925$ were considered. 
In that region of masses the observed effect coming from 
the dynamics of the gluinos was mainly the overall renormalization 
of the lattice spacing due to the fermionic virtual loops.
The change of dimensionless ratios of masses and string tension
where only moderate up to $\kappa\!\leq\!0.1925$ where most
of the simulations were performed.

The set of configurations for the lightest gluino produced
in \cite{CaetAl}, $\kappa$=0.1925, is taken here as a starting point. 
We further simulate the model at
lighter gluinos, at $\kappa\!=\!0.194$ and 0.1955.
The largest of these hopping parameters coincides with
the central value of the estimate of $\kappa_c$ 
from the study of the finite volume gluino condensate,
$\kappa_c\!=\!0.1955(5)$ \cite{KietAl2}.
That determination was however obtained 
on a relatively small lattice ($6^3\times 12$) and the
value of $\kappa_c$ is likely to be underestimated.
In fact, anticipating results of the present study, 
the gluino mass starts decreasing significantly
only for $\kappa\gtrsim 0.194$. For $\kappa\!=\!0.1955$
the gluino mass is quite small but still appreciably different from zero.
This is also evident in the simulation process.

Simulating light fermions in a reasonably large physical volume 
is a challenging task from the algorithmic point of view. 
The difficulty is related to very small eigenvalues
of the fermion  matrix. 
The relevant parameter in this context is the condition number 
of $\tilde{Q}^2$, where $\tilde{Q}\equiv\gamma_5 Q$
is the Hermitian fermion matrix\footnote{The condition number 
of a matrix is defined as the ratio between its largest and  smallest eigenvalue.}.
For a given simulation volume the condition number gives 
an indication of the `lightness' of the gluino. 
For the lightest gluino, at $\kappa\!=\!0.1955$,
we had condition numbers ${\cal O}(10^5)$.
A direct comparison with the more familiar case of QCD
is not possible since the simulation of SYM is generally less demanding. 

Another difficulty related to light fermions
is the shrinking of the physical volume due to the 
renormalization of the lattice spacing by fermionic virtual loops.
We expect that already at $\kappa$=0.194 the low energy 
bound-states mass spectrum is strongly affected by the finite-size scaling
on our $12^3\times24$ lattice. (See also Subsec.~\ref{sec:r0}).
The situation is different for the main subject of this work, 
the SUSY WIs. The WIs hold also on a finite volume with volume-dependent coefficients.
These are however essentially renormalizations defined at the scale of the UV cutoff $a^{-1}$.
Our volumes should be consequently large enough for an accurate determination.

All the numerical computations of this work were performed on the
two 512-nodes CRAY-T3E machines at the John von Neumann Institute for Computing (NIC),
J\"ulich, with 307.2 and 614.4 GFLOPS peak-performance respectively.
The CPU cost of the simulation was $\sim$1 GFLOPS Year sustained
($\sim$$3\cdot 10^{16}$ f.p.o.)
for each of the two simulation points of this work.

\subsection{The TSMB algorithm: simulation parameters}

\begin{table}[t]
\begin{center}
\parbox{15cm}{\caption{\label{runs_tab}
 Parameters of the numerical simulations at $\beta=2.3$.
 The run at $\kappa\!=\!0.1925$ was performed in \cite{CaetAl}.
 The notation is explained in the text.
}}
\end{center}
\begin{center}

\begin{tabular}{llcccccccccccc }
\hline
\multicolumn{1}{c}{$\kappa$}  &
\multicolumn{1}{c}{$\epsilon\cdot 10^4$}  &
$\lambda$  & $n_1$  &  $n_2$  &  $n_3$  &  $n_4$  &
$n_{HB}$ & $n_{OB}$ & $n_M$ & $n_{NC}$ &
\multicolumn{1}{c}{updates}  &  \multicolumn{1}{c}{offset} & $N_{lat}$  
\\
\hline
0.1925 &  $3.0$ &  3.7  &  32  &  150  &  220  &  400  &
1 & 3 & 1 & 1 &
216000  & 50 & 9  \\

0.194 (a)&  $0.8$ &  4.5  &  38  &  280  &  320  &  400  &
1 & 2 & 8 & 8 &
2250  & 50 & 9  \\

0.194 (b) &  $1.0$ &  4.5  &  24  &  160  &  200  &  400  &
1 & 2 & 8 & 8 &
2700  & 20 & 9  \\

0.194 (c)&  $1.0$  &  4.5  &  24  &  120  &  160  &  400  &
1 & 2 & 8 & 8 &
1620  & 20 & 9  \\

0.194 (d)&  $1.0$ &  4.5  &  28  &  160  &  210  &  400  &
2 & 2 & 8 & 8 &
35460 &  20 & 9  \\

0.1955 (a)&  $0.2$ & 5.0  &  32  &  420  &  560  &  840  &
2 & 10 & 4 & 4 &
5040 &  30 & 8  \\

0.1955 (b)&  $0.125$ &  5.0  &  32  &  480  &  640  &  960  &
2 & 10 & 4 & 4 &
27672  & 15 & 8  \\

0.1955 (c)&  $0.125$ &  5.0  &  32  &  480  &  640  &  960  &
6 & 6 & 12 & 12 &
33120  &  10 & 8  \\

\hline
\end{tabular}
\end{center}
\end{table}

The TSMB algorithm used in the simulations is
defined in \cite{Mont3,Mont4}. 
The multi-bosonic updating with the scalar pseudofermion fields 
was performed by heatbath and overrelaxation 
for the scalar fields and Metropolis sweeps for the gauge field;
we refer to \cite{CaetAl} for more details on the implementation 
of the TSMB algorithm in the case of the N=1 SU(2) SYM.

In Table~\ref{runs_tab} we report the parameters for the simulations
performed in this work. We also include for reference the
parameters for the simulation at $\kappa\!=\!0.1925$
performed in~\cite{CaetAl}. 
We briefly explain the meaning of the various symbols
(see also~\cite{CaetAl}); \newline\noindent
$[\epsilon,\, \lambda]$ (columns 2 and 3)
is the presumed domain of the eigenvalue spectrum of $\tilde{Q}^2$;
this is also the domain of validity of the polynomial approximation
of the fermionic measure;
$n_{1,\ldots,4}$ (columns 4 to 7) are the orders of the polynomial 
approximations used in the simulation and measurement process;
in particular, $n_{1}$ and $n_{2}$ are the orders of the 
polynomial approximations 
in the local update and in the global accept-reject step (noisy correction)
respectively; $n_{3}$ is the order of the polynomial used for the 
generation of the noisy vector in the noisy correction; $n_4$ is the order 
of the polynomial used for the computation of the reweighting factors
in the measurements; $n_{HB}, n_{OB}, n_{M}$ (columns 8 to 10)
indicate the heatbath, overrelaxation and Metropolis sweeps performed at each step
of the local update; $n_{NC}$ (column 11) is the number of 
Metropolis sweeps separating two consecutive global accept-reject 
steps; finally columns 12 to 14 report 
the total number of updates at equilibrium, the offset between measurements,
and the number of independent lattices simulated.

\begin{table}[b]
\begin{center}
\parbox{15cm}{\caption{\label{autocorr_tab}
 Exponential and integrated autocorrelation of the plaquette
 measured in update cycles. In square brackets the integrated autocorrelation
 is measured in no. of Dirac matrix multiplications.
 The data at $\kappa\!=\!0.1925$ are taken from \cite{CaetAl}.
 }}
\end{center}

\begin{center}
\begin{tabular}{lllll}
\hline
\multicolumn{1}{c}{$\kappa$ }  &
\multicolumn{1}{c}{run}  &
\multicolumn{1}{c}{$\tau_{exp}$}  &
\multicolumn{1}{c}{$\tau_{int}$}  \\
\hline

0.1925 & \multicolumn{1}{c}{-} & 378(37) & 675(200)  [1.15(34) $10^6$]  \\
0.194  & \multicolumn{1}{c}{d} & 249(68) & 272(83)\z\   [0.75(23) $10^6$]  \\
0.1955 & \multicolumn{1}{c}{b} & 220(50) & 280(70)\z\   [1.49(37) $10^6$]  \\
0.1955 & \multicolumn{1}{c}{c} & 210(40) & 250(40)\z\   [1.71(27) $10^6$]  \\
0.1955 & \multicolumn{1}{c}{b,c} & 260(30) & 420(50)   \\

\hline
\end{tabular}
\end{center}
\end{table}

\subsection{Autocorrelations}

Tuning the various parameters of the algorithm
($\epsilon$, $\lambda$, $n_{1,2,3}$, $n_{HB}$, $n_{OV}$, etc.)
is essential to get an optimized updating. As an optimization criterion 
we have considered the autocorrelation of the plaquette.
In particular, the order of the first polynomial 
$n_1$ was increased until the acceptance probability 
in the noisy correction reached $\sim$50$\%$. With this choice 
the two steps of the updating process equally contribute
in shaping the distribution of gauge configurations.
Larger values of $n_1$ have the effect of increasing autocorrelations
with no substantial improvement of the algorithm.
In the runs at $\kappa\!=\!0.1955$ the update of the pseudofermionic
fields was performed by iterating twice a sub-sequence 
of $n_{HB}/2$ heatbath and $n_{OB}/2$ overrelaxation sweeps.
In Table~\ref{autocorr_tab} the integrated autocorrelations
for the various runs are reported. The data for $\kappa\!=\!0.1925$
are taken from \cite{CaetAl}. 
A better tuning of the parameters of the algorithm allowed to keep 
autocorrelations down at low levels in spite of an increasingly 
light gluino. 

\subsection{Smallest eigenvalues and reweighting factors}

\begin{figure}[t]
\begin{center}
\leavevmode
\epsfig{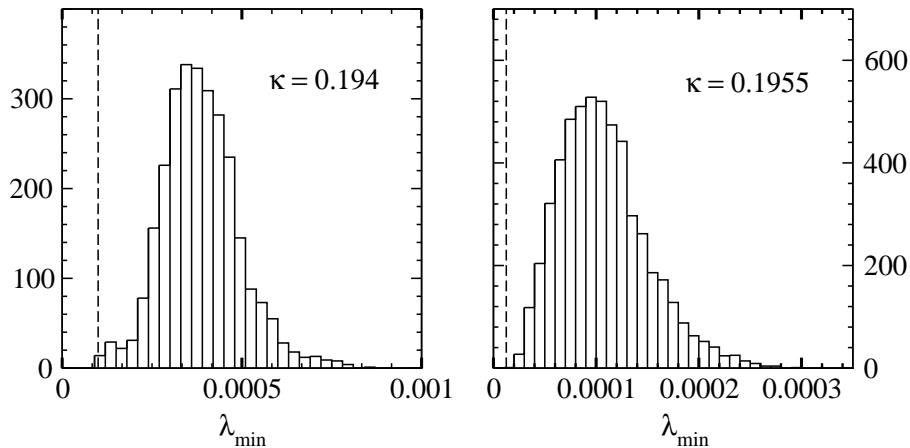}\\
\end{center}
\caption{\label{fig:minev} Distribution of the smallest eigenvalue 
of $\tilde{Q}^2$ for the two simulation points of this work.
The dashed line indicates the value of $\epsilon$ used in the 
simulation.}
\end{figure}

To monitor the accuracy of the polynomial approximation
in the updating process, we constantly checked the smallest and largest 
eigenvalue of $\tilde{Q}^2$. The distribution of the smallest eigenvalue
for the two simulation points of this work is reported in
Fig.~\ref{fig:minev}. The vertical dashed line indicates the
value of $\epsilon$ used in the simulation.

\begin{figure}[t]
\begin{center}
\leavevmode
\epsfig{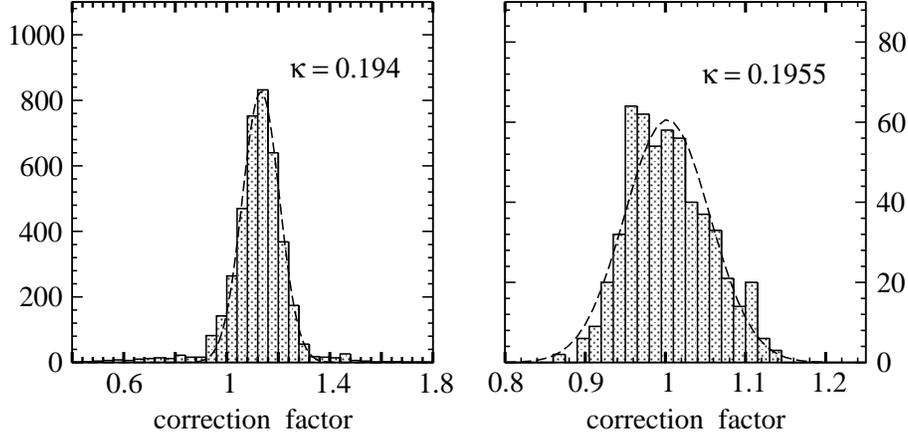}\\
\end{center}
\caption{\label{fig:corr} Distribution of the reweighting factors with gaussian fit.}
\end{figure}

Extremely small eigenvalues can exceptionally occur 
without substantial harm.
The corresponding configuration would then be suppressed 
at the measurement level by the reweighting.
We calculated reweighting factors for sub-samples 
of configurations.
These turn out to be gaussian distributed with average $\sim\!1$, 
see Fig.~\ref{fig:corr}. 
For $\kappa\!=\!0.194$ we also observe a short tail towards small values. 
These  distributions are
consistent with the absence of extremely small 
eigenvalues in the ensembles. The effect of the reweighting 
turns out to be negligible compared to the statistical fluctuations
for the quantities considered in this study.
This confirms the overall accuracy of the simulation algorithm.
 
\subsection{The sign of the Pfaffian} 

The fermionic measure implemented in the updating algorithm
is given by $\sqrt{\det(Q)}$. The actual measure
for Majorana fermions is instead given by the Pfaffian 
of the anti-symmetric matrix $M=CQ$
\be
{\rm Pf}(M)=\sqrt{\det(Q)}\cdot {\rm sign}({\rm Pf}(M))\ .
\ee
The previous formula implies that the configurations
obtained by Monte Carlo updating should be further reweighted in the measurements
by ${\rm sign}({\rm Pf}(M))$ .
This could potentially introduce difficulties. 
Indeed, were positive and negative signs almost equally distributed in 
our ensembles, a cancellation could occur in the
statistical averages. The significance of the samples
would then be close to zero. This occurrence is know as `the sign problem'.
In the N=1 SYM with Wilson fermions the two signs are expected 
to be equally distributed 
for $\kappa>\kappa_c$. For $\kappa<\kappa_c$,
sign-flips should be suppressed if the volume is large enough
(the situation right at the critical point is still unclear).

\begin{figure}[t]
\begin{center}
\leavevmode
\epsfig{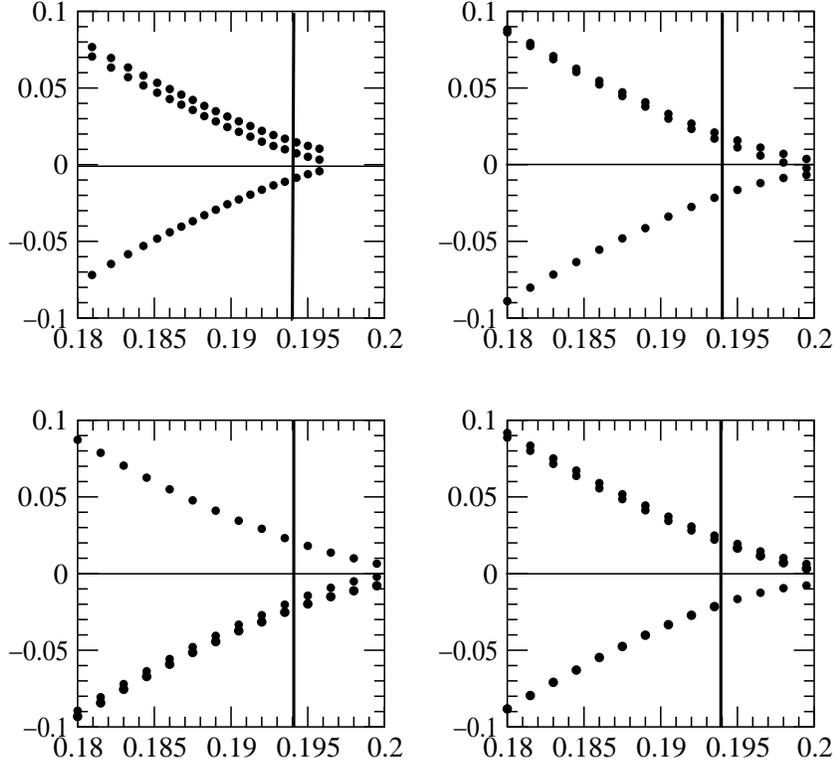}\\
\end{center}
\caption{\label{fig:flow} The flow of eigenvalues as a function of the hopping parameter
for different configurations produced at $\beta=2.3$ and $\kappa=0.194$.
The vertical line indicates the hopping parameter of the simulation.}
\end{figure}

We determined the sign of the Pfaffian for
a subsample (10$\%$) of configurations at $\kappa$=0.194
on our $12^3\times 24$ lattice.
We used the method explained in \cite{CaetAl} consisting in following
the flow of the eigenvalues of $\tilde{Q}$ as a function
of the hopping parameter. 
A theorem ensures ${\rm sign}({\rm Pf}(M))\!=\!1$ for small $\kappa$'s. 
The sign of the Pfaffian flips whenever an eigenvalue crosses zero
in the flow. We never found such zero-level crossings, implying 
that ${\rm sign}({\rm Pf}(M))$ is always positive 
for the considered sub-sample. 
Examples of eigenvalue flows are given in Fig.~\ref{fig:flow}.
We conclude that statistically less than 0.5$\%$ of the configurations 
have negative sign.
Negative Pfaffians were detected in~\cite{CaetAl} on a smaller lattice ($6^3\times 12$) 
for $\kappa\geq 0.196$.
Absence of sign-flip of the Pfaffian in our ensembles is also 
supported by the observation that extremely small eigenvalues 
do not occur and, consistently, the distribution of the 
reweighting factors does not extend to zero. Indeed each sign-flip under continuous
modification of the gauge configuration would imply 
the crossing of a configuration with an exact fermionic zero-mode.

\subsection{Determination of $r_0/a$}
\label{sec:r0}

\begin{table}[t]
\begin{center}
\parbox{15cm}{\caption{\label{tab:r0}
 Determination of $r_0/a$ for the sets of configurations considered in this study.
 We also report the ratio $L_x/r_0$ where $L_x$ is the spatial lattice size.}}
\end{center}

\begin{center}
\begin{tabular}{lll}
\hline
\multicolumn{1}{c}{$\kappa$ }  &
\multicolumn{1}{c}{$r_0/a$}  &
\multicolumn{1}{c}{$L_x/r_0$}  \\
\hline
0.1925 & 6.71(19) & 1.79(5)  \\
0.194 & 7.37(30) & 1.63(7)  \\
0.1955 & 7.98(48) & 1.50(9)  \\
\hline
\end{tabular}
\end{center}
\end{table}

In \cite{Som94} the scale parameter $r_0$ has been proposed as a
reference scale for gauge configurations. It is defined by
\be
r_0^2F(r_0)=1.65
\ee
where $F(r)$ is the force between static fermionic color sources
in the fundamental representation.
In our case the determination of $r_0/a$ allows us to monitor 
the reduction of the simulation volume with increasing $\kappa$.

We first calculated the static potential between color sources
in the fundamental representation $V({\bf r})$;
this is needed only at intermediate distances.
The potential can be estimated from Wilson loops $W({\bf r},t)$ by
\begin{equation}
V({\bf r},t)= \ln \left( \frac{W({\bf r},t)}{W({\bf r},t+1)}\right)\ .
\end{equation}
To reduce the noise we used APE smearing \cite{Alb87} on the spatial links with
$n_{APE}=12$ iterations and $\epsilon_{APE}=0.5$. Nevertheless the results get unstable
for the larger Wilson loops, so we could not make an extrapolation to large
$t$. We instead took the result for $t=2$, calculated $r_0/a$ for these
values and added the difference to the result from $t=3$ to the total error.
 
To get $r_0/a$ from the potential we followed
\cite{Edw98,All01}. Their general procedure would be
to make a fit to
\begin{equation}
V({\bf r})=V_0+\sigma r-e\left[\frac{1}{\bf r}\right]+f\left(
\left[\frac{1}{\bf r}\right]-\frac{1}{r}\right)
\end{equation}
where
\begin{equation}
\left[\frac{1}{\bf r}\right]
=4\pi \int_{-\pi}^{\pi}\frac{\mbox{d}^3k}{(2\pi)^3}
\frac{cos({\bf k} \cdot {\bf r})}{4\sum_{j=1}^3\sin^2(k_j/2)}
\end{equation}
removes lattice artifacts exactly to lowest order of perturbation theory.
However it was not possible with our data to make fits to all four parameters.
We therefore followed \cite{Edw98} and made a three parameter fit
fixing $e=\pi/12$ and a two parameter fit fixing in addition $f=0$.
One can now extract $r_0$ as
\begin{equation}
r_0=\sqrt{\frac{1.65-e}{\sigma}}\ .
\end{equation}
Our final estimate for $r_0$ comes from the three parameter fit, and the
difference to the two parameter fit was again added to the Jackknife
estimate for the error.       

In Table~\ref{tab:r0} we report the results for $r_0/a$ for the  
three sets of configurations considered in this study. 
From data in column 3 we see that the reduction of the physical 
lattice size is less than $20\%$ for the lightest gluino in our simulations.

\section{Numerical analysis of the SUSY WIs}

In Section~\ref{sec:operators} we have shown how the quantities $am_S Z_S^{-1}$ and  
$Z_T Z_S^{-1}$ can be obtained by solving the system of equations~(\ref{system})
for a given insertion operator. We have argued that there are two 
such independent dimension-$7/2$ operators. 
We choose the operators $\chi^{(sp)}(x)$ and $T^{(loc)}_0(x)$ 
defined in~(\ref{phi1}) and~(\ref{T2}) for our analysis.
Discrepancies between results obtained with these two operators
signal the presence of systematic effects due to the lattice
discretization.

In practice we use two methods to obtain $am_S Z_S^{-1}$ and $Z_T Z_S^{-1}$.
The first method is the most straightforward, consisting simply in
solving the system of equations~(\ref{system}) for each time-separation~$t$. 
Results obtained in this way are reported in Figs.~\ref{fig:mass} and~\ref{fig:zeta}; 
they should be independent of $t$ when contact terms are absent
for large enough time-separations.
The second method, explained in detail in
Appendix~D, consists in constructing an overdetermined system of equations for
several consecutive time-separations $(t_{min},\cdots, L_t/2$) and fitting simultaneously
for all time-separations. Results obtained in this way are reported in 
Tables~\ref{tab:mass_ps}-\ref{tab:zeta_loc}.
Care should be taken in choosing $t_{min}$ so as to avoid time-separations 
in which sink and insertion operators give rise to contact terms.

Previous experience with the mass spectrum \cite{CaetAl} shows that point-like 
projection-operators give a poor signal for the correlations.
This is mainly due to the gluonic content of both sink and insertion operators.
The inconvenience is expected to be even more severe
in the present case. Indeed
sink operators involve time-derivatives  which are subject to large 
statistical fluctuations.
The problem was solved in \cite{CaetAl,DoetAl} by smearing the projection-operator
for the gluino-glue bound state.
Combined APE~\cite{Alb87} and Jacobi~\cite{jacobi} smearing were performed 
on the gluon and gluino fields respectively.
In the present case we apply the same procedure only
for the insertion operator.
We searched for the optimal smearing parameters by analyzing
sub-samples of gauge configurations.
The set of parameters employed in the final analysis was:
$N_{Jacobi}$=18, $K_{Jacobi}$=0.2, $N_{APE}$=9, $\epsilon_{APE}$=0.5 (set~A).
In one case ($\kappa=0.1925$ and insertion operator $\chi^{(sp)}(x)$)
we have also considered $\epsilon_{APE}$=0.1 and remaining parameters as in set~A
(set~B).
The APE-Jacobi procedure is not completely satisfactory in the
case of insertion operators extended in the time-direction such as $T^{(loc)}_0(x)$.
In our case with dynamical gluinos,
a multi-hit procedure~\cite{PaPeRa83} on the temporal links
would not work since the noisy correction would be ineffective and 
anyway far too expensive.

The inversions of the fermion matrix $Q$ required for the computation 
of the correlations were performed by the conjugate-gradient method.
The number of iterations necessary for a good accuracy on the final result
increases for light gluinos. Convergence was
improved by preconditioning the Hermitian fermion matrix.
The residuum for the conjugate-gradient
was chosen by requiring that the final accuracy on the determination 
of $am_S Z_S^{-1}$ and  $Z_T Z_S^{-1}$ was $\lesssim\! 5\cdot 10^{-5}$. With this choice,
$\sim\! 1100$ iterations were needed on average for the 
lightest gluino at $\kappa=0.1955$. The computing power used was correspondingly 
$\sim\! 6\cdot 10^{15}$ f.p.o.; this is about 20$\%$ of the amount
employed for the generation of the gauge-fields.
The site $y$ of the insertion was chosen randomly for each configuration.
We checked correlations in simulation time between propagators involved in the WIs.
With the random choice of $y$ the correlations 
between two consecutively measured propagators turns out
to be negligible (less than 0.05).
Consequently a naive jackknife procedure can be used for the error analysis
on $am_S Z_S^{-1}$ and  $Z_T Z_S^{-1}$.

At $\kappa=0.1925$ we have also used 
a version of the operator $\chi^{(sp)}(x)$ defined as in Eq.~(\ref{phi1}) but
with the lattice field tensor given by the simple plaquette
\be\label{plaquette}
P^{(pl)}_{\mu\nu}(x) = \frac{1}{2ig_0a^2}
\left(U_{\mu\nu}(x)-U^{\dagger}_{\mu\nu}(x)\right)\ .
\ee
The drawback of this definition is that the properties of transformation 
under \ppp\ and \ttt\ are not the same as in  the continuum. 
However comparison of results obtained with the two definitions of 
$\chi^{(sp)}(x)$ gives an indication of the size of discretization
errors. 

\subsection{Results}

\begin{figure}[t]
\begin{center}
\leavevmode
\epsfig{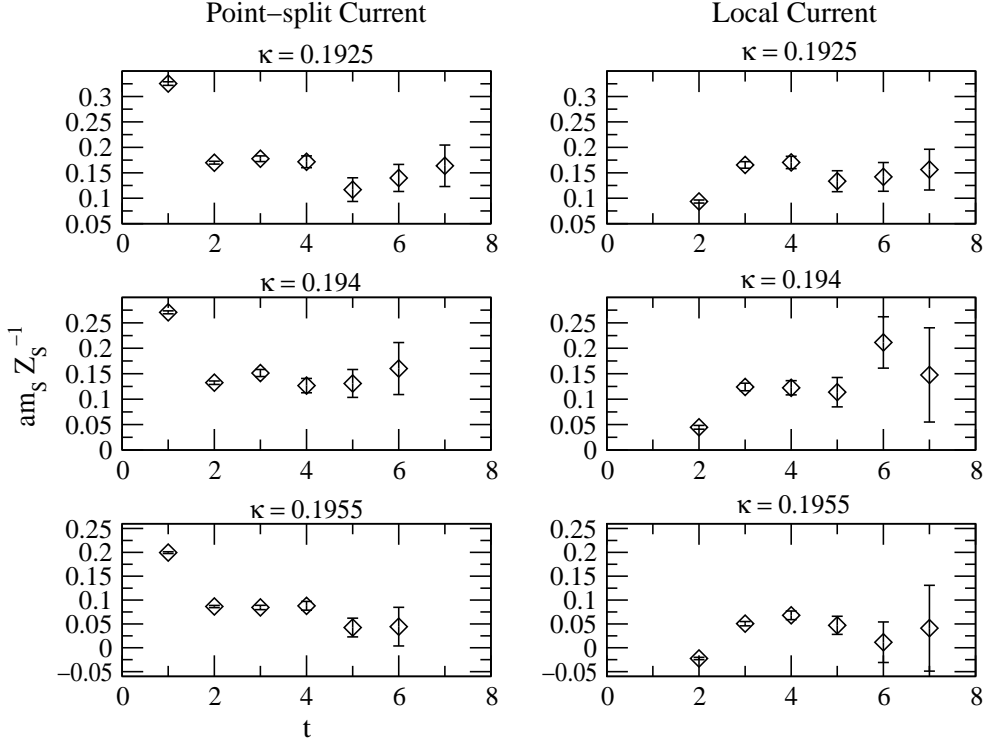}\\
\end{center}
\caption{\label{fig:mass} $am_S Z_S^{-1}$
as a function of the time-separation $t$ with insertion operator $\chi^{(sp)}(x)$.}
\end{figure}

\begin{figure}[t]
\begin{center}
\leavevmode
\epsfig{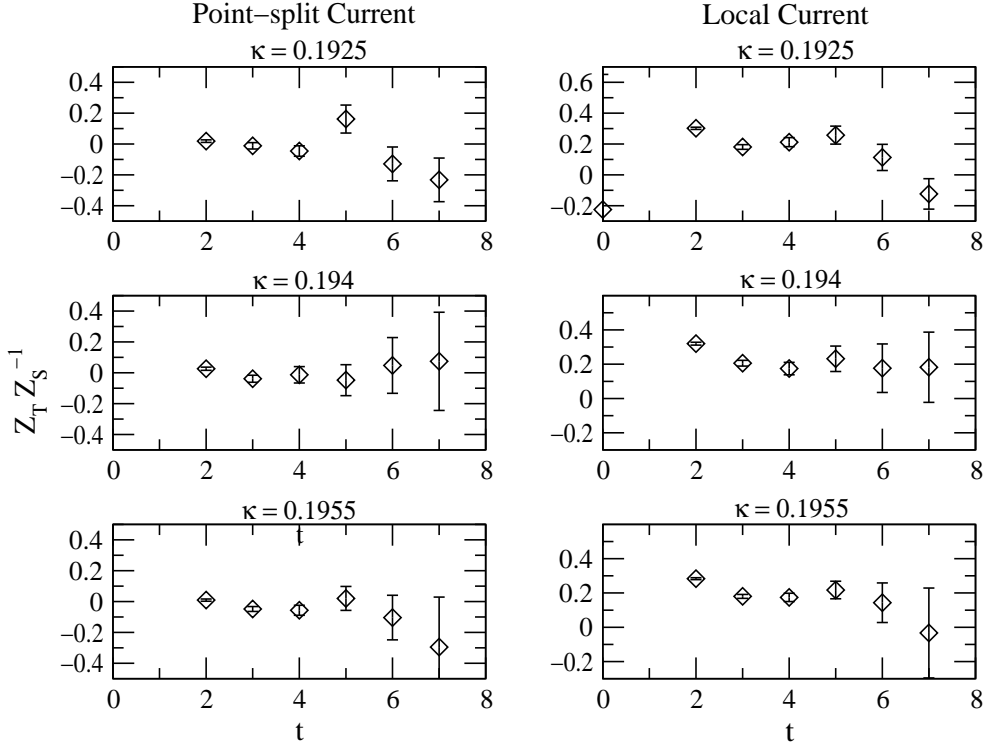}\\
\end{center}
\caption{\label{fig:zeta} $Z_T Z_S^{-1}$ as a function of the time-separation $t$
with insertion operator $\chi^{(sp)}(x)$.}
\end{figure}

In Figs.~\ref{fig:mass} and \ref{fig:zeta} we report 
the determinations of $am_S Z_S^{-1}$ and $Z_T Z_S^{-1}$ 
respectively, as a function of the time-separation $t$.
The left column refers to the point-split SUSY current $S_\mu^{(ps)}(x)$ 
while the right one to the local current $S_\mu^{(loc)}(x)$.
The insertion operator is $\chi^{(sp)}(x)$, which generally
gives the best signal.
A behavior consistent with a plateau can be observed for time-separations
$t\geq 3$ for which there is no contamination from contact terms.
The signal is rapidly washed-out by the statistical fluctuations
for large time-separations.

The insertion operator $T^{(loc)}_0(x)$ containing links in the time direction 
gives larger fluctuations than the time-slice operator $\chi^{(sp)}(x)$. 
This is probably related to the poor performance of the 
Jacobi-APE smearing on operators extended in the time direction.

For a given insertion operator, $Z_T Z_S^{-1}$  is subject 
to larger statistical fluctuations than  $am_S Z_S^{-1}$ 
as can be seen from Fig.~\ref{fig:zeta} (notice the different scale). 
We have no theoretical justification for this outcome.

The results for $am_S Z_S^{-1}$ and $Z_T Z_S^{-1}$ from a global 
fit over a range of time-separations $t\geq t_{\rm min}$
are reported in Tables~\ref{tab:mass_ps}-\ref{tab:zeta_loc}.
Data are obtained by solving the overdetermined linear system 
as explained in Appendix~D. An equivalent procedure consists in performing
a least mean square fit on $am_S Z_S^{-1}(t)$ and $Z_T Z_S^{-1}(t)$
for $t\geq t_{\rm min}$ taking time correlations into account.
This second procedure gives results consistent with the first one, 
with $\chi^2/d.o.f.\approx 1$ when contact terms are absent.
For the final estimates we take $t_{\rm min}=3$ for time-slice operators and 
$t_{\rm min}=4$ for operators extended in the time-direction.
This choice ensures absence of contact terms.

Discretization effects can be checked by comparing determinations
obtained with the two independent insertions 
$\chi^{(sp)}(x)$ and $T^{(loc)}_0(x)$. For $\kappa=0.1925$ one can also 
compare between different definitions of $\chi^{(sp)}(x)$
(simple plaquette definition of the lattice field tensor, 
Eq.~(\ref{plaquette}), and different smearing parameters).
It should be recalled at this point that different discretizations 
of the currents give different values for $Z_S$ and $Z_T$.
Consequently data from different discretizations should not be confronted.

\begin{figure}[t]
\begin{center}
\leavevmode
\epsfig{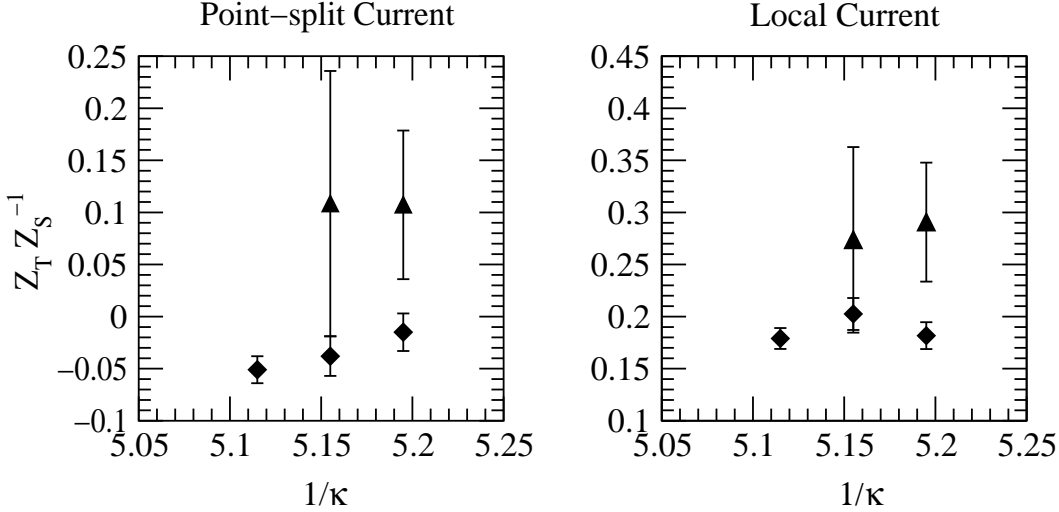}\\
\end{center}
\caption{\label{fig:zeta_k} $Z_T Z_S^{-1}$ as a function of $1/\kappa$
with the insertion operator $\chi^{(sp)}(x)$ (filled diamonds) and  $T_0^{(loc)}(x)$
(filled triangles).}
\end{figure}

\begin{figure}[t]
\begin{center}
\leavevmode
\epsfig{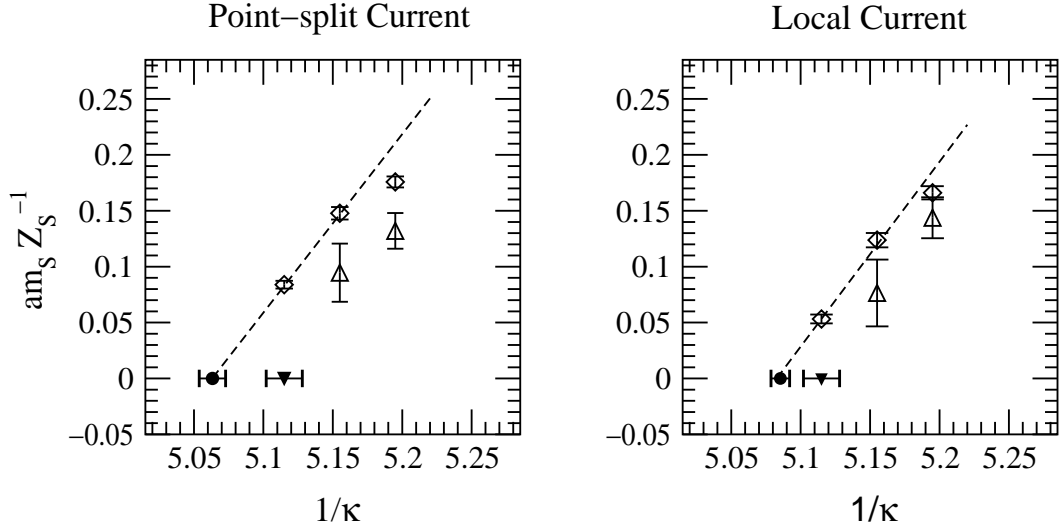}\\
\end{center}
\caption{\label{fig:extra_mass} $am_S Z_S^{-1}$ as a function of $1/\kappa$
with the insertion operator $\chi^{(sp)}(x)$ (diamonds) and  $T_0^{(loc)}(x)$
(triangles). A linear extrapolation is also reported. The filled triangle indicates 
the determination of $\kappa_c$
from the first order phase transition of \cite{KietAl}.}
\end{figure}

An interesting point is the dependence of $Z_T Z_S^{-1}$
and $am_S Z_S^{-1}$  on the hopping parameter $\kappa$. 
This is reported in Fig.~\ref{fig:zeta_k} and~\ref{fig:extra_mass}
respectively. Data refer to determinations obtained with $t_{min}=3$
for insertion operator $\chi^{(sp)}(x)$ and 
$t_{min}=4$ for insertion operator $T_0^{(loc)}(x)$.
The definition of $\chi^{(sp)}(x)$ is the one with 
clover field tensor and smearing parameters of set~A
(cf. Tables~\ref{tab:mass_ps}-\ref{tab:zeta_loc}).
We see that the combination of renormalization factors $Z_T Z_S^{-1}$   
shows no appreciable dependence on $\kappa$.
We recall (see discussion in Subsec.~\ref{sec:renorm})
that the latter is an $O(a)$ effect.
Fitting these results (only for insertion $\chi^{(sp)}(x)$) 
with a constant in $1/\kappa$
we obtain $Z_T Z_S^{-1}=-0.039(7)$, for the point-split current, and  
$Z_T Z_S^{-1}=0.185(7)$ for the local current.
The renormalization is surprisingly small in the case of 
the point-split current. 
An estimate of $Z_T Z_S^{-1}$ for the point-split current at
$\beta=2.3$ can be obtained from the 1-loop perturbative 
calculation in \cite{Tani}. At order $g_0^2$ one obtains 
$Z_T Z_S^{-1}\equiv\left.Z_T\right|_{\rm 1-loop}=-0.074$.

In  Fig.~\ref{fig:extra_mass} the determination of $am_S Z_S^{-1}$ 
is reported as a function of the inverse hopping parameter.
The expectation is that $am_SZ_S^{-1}$ vanishes linearly when 
$\kappa\rightarrow\kappa_c$. 
We see a clear decrease when $\kappa$ is increased towards $\kappa_c$;
$am_S Z_S^{-1}$ starts dropping  abruptly
at $\kappa\gtrsim 0.194$. We refer to data with insertion $\chi^{(sp)}(x)$.
Given the weak dependence of 
the renormalization factors on the hopping parameter, 
the relative decrease of $am_S Z_S^{-1}$ as a function of $\kappa$ 
should roughly compare with that of the gluino mass itself. 
In the case of the point-split current 
the latter almost halves when passing from 
$\kappa\!=\!0.194$ to $\kappa\!=\!0.1955$.
In the case of the local current we get an even smaller mass
at $\kappa\!=\!0.1955$. 
Finally we obtain a determination $\kappa_c$ by performing 
an extrapolation to zero gluino mass from two 
largest $\kappa$-values. The result is
$\kappa_c=0.19750(38)$ with the point-split current
and $\kappa_c=0.19647(27)$ with the local one.
These values can be compared with the previous determination
from the phase transition $\kappa_c=0.1955(5)$ \cite{KietAl}. 

\section{Summary and conclusions}

The present study shows that the extraction of the ratios
$am_S Z_S^{-1}$ and $Z_T Z_S^{-1}$ from the on-shell SUSY Ward identities
is technically feasible with the computing resources at hand.
The main technical difficulty (related to SUSY) is that 
high-dimensional operators with a mixed gluonic-fermionic composition 
must be considered, introducing relatively large statistical fluctuations.
This difficulty can be handled with an appropriate smearing
procedure. The non-perturbative determination of the
ratio $am_S Z_S^{-1}$ can be used for 
a determination of the critical hopping parameter 
$\kappa_c$ corresponding to massless gluinos. This can be compared
to the independent determination of $\kappa_c$, based on
the chiral phase transition~\cite{KietAl}.
The fact that the two determinations are not in full agreement
can be explained by the presence of systematic effects.
These are predominantly $O(a)$ effects in our case and
finite volume effects in the determination of Ref.~\cite{KietAl}.
The discretization error in the present approach can be checked by 
comparing results from the two independent insertions.
The statistical uncertainty is however relatively large (10-40$\%$)
for the insertion operator involving links in the time-direction.
Our results for $am_S Z_S^{-1}$ and $Z_T Z_S^{-1}$ are consistent 
with the WIs~(\ref{renormward}) with $O(a)$
effects comparable to the statistical errors.
The non-perturbative results here presented will be complemented by 
an analytical perturbative calculation \cite{pert}.

The overall conclusion of our study is that lattice SUSY WIs
can be implemented in the non-perturbative determination of the
correctly subtracted SUSY current and provide a practicable method
for the verification of SUSY restoration in this framework,
once the proper SUSY current is identified and the usual lattice
artefacts (most notably finite cutoff  effects) are kept under control.


\vspace*{1em}
 {\bf Acknowledgement:}
 The numerical study presented here has been performed on the
 CRAY-T3E computers at the John von Neumann Institute for Computing (NIC),
 J\"ulich.
 We thank NIC and the staff at ZAM for their kind support.
 A.V. and C.G. wish to thank the DESY theory group and 
 the INFN-Rome 2 respectively for their hospitality.


\newpage

\begin{table*}[t]
\begin{center}
\parbox{15cm}{\caption{\label{tab:mass_ps}
Summary of the results for $am_S Z_S^{-1}$ at $\beta=2.3$
with point-split currents.}}
\begin{tabular}{@{}lclll}
\hline
\multicolumn{1}{c}{$\kappa$} & operator & $t_{min}=3$ & $t_{min}=4$ &
$t_{\rm min}=5$ \\
\hline
0.1925 & $\chi^{(sp)}$      & 0.176(5)   & 0.166(10) & 0.135(14)     \\
0.1925 & $\chi^{(sp)}$ (*)  & 0.182(6)   & 0.152(11) & 0.150(16)     \\
0.1925 & $\chi^{(sp)}$ (**) & 0.1969(47) & 0.168(9)  & 0.136(14)     \\
0.1925 & $T^{(loc)}_0$             &            & 0.132(16) & 0.124(21)     \\ 
\hline
0.194  & $\chi^{(sp)}$      & 0.148(6)   & 0.130(11) & 0.146(21)     \\  
0.194  & $T^{(loc)}_0$             &            & 0.095(27) & 0.090(27)     \\ 
\hline
0.1955 & $\chi^{(sp)}$      & 0.0839(35) &  0.0820(7) & 0.053(14)      \\
\hline
\end{tabular} \\
\parbox{11cm}{* \footnotesize With plaquette field tensor.}\\
\parbox{11cm}{** \footnotesize With plaquette field tensor and smearing parameters B.}\\
\end{center}
\vspace{1.5cm}
\end{table*}

\begin{table*}[htb]
\begin{center}
\parbox{15cm}{\caption{\label{tab:mass_loc}
Summary of the results for $am_S Z_S^{-1}$ at $\beta=2.3$
with local currents.}}
\begin{tabular}{@{}lclll}
\hline
\multicolumn{1}{c}{$\kappa$} & operator & $t_{min}=3$ & $t_{min}=4$ &
$t_{\rm min}=5$ \\
\hline
0.1925 & $\chi^{(sp)}$      & 0.166(6)  & 0.166(11) & 0.146(16)     \\
0.1925 & $\chi^{(sp)}$ (*)  & 0.173(6)  & 0.155(11) & 0.135(19)     \\
0.1925 & $\chi^{(sp)}$ (**) & 0.1821(47)  & 0.173(11) & 0.154(19)     \\
0.1925 & $T^{(loc)}_0$             &           & 0.144(18) & 0.143(25)     \\ 
\hline
0.194  & $\chi^{(sp)}$      & 0.124(6)  & 0.126(12) & 0.142(24)     \\  
0.194  & $T^{(loc)}_0$             &           & 0.076(30) & 0.098(35)     \\ 
\hline
0.1955 & $\chi^{(sp)}$      & 0.0532(40)  & 0.064(8)  & 0.047(15)      \\
\hline
\end{tabular} \\
\parbox{11cm}{* \footnotesize With plaquette field tensor.}\\
\parbox{11cm}{** \footnotesize With plaquette field tensor and smearing parameters B.}\\
\end{center}
\end{table*}

\newpage

\begin{table*}[t]
\begin{center}
\parbox{15cm}{\caption{\label{tab:zeta_ps}
Summary of the results for $Z_T Z_S^{-1}$ at $\beta=2.3$ 
with point-split currents.}}
\begin{tabular}{@{}lclll}
\hline
\multicolumn{1}{c}{$\kappa$} & operator & $t_{min}=3$ & $t_{min}=4$ &
$t_{\rm min}=5$ \\
\hline
0.1925 & $\chi^{(sp)}$      &$-$0.015(19)  &$-$0.036(31)  &\m 0.045(56)     \\
0.1925 & $\chi^{(sp)}$ (*)  &$-$0.044(16)  &$-$0.096(33)  &\m 0.01(6)       \\
0.1925 & $\chi^{(sp)}$ (**) &$-$0.058(14)  &$-$0.044(32)  &$-$0.07(5)       \\
0.1925 & $T^{(loc)}_0$             &              &\m 0.11(7)    &$-$0.03(7)       \\ 
\hline
0.194  & $\chi^{(sp)}$      &$-$0.038(19)   &$-$0.024(43)  &$-$0.08(7)        \\  
0.194  & $T^{(loc)}_0$             &               &\m 0.11(13)   &\m 0.02(13)       \\ 
\hline
0.1955 & $\chi^{(sp)}$      &$-$0.051(13)   &$-$0.064(26)  &$-$0.05(5)        \\
\hline
\end{tabular} \\
\parbox{11cm}{* \footnotesize With plaquette field tensor.}\\
\parbox{11cm}{** \footnotesize With plaquette field tensor and smearing parameters B.}\\
\end{center}
\vspace{1.5cm}
\end{table*}

\begin{table*}[htb]
\begin{center}
\parbox{15cm}{\caption{\label{tab:zeta_loc}
Summary of the results for $Z_T Z_S^{-1}$ at $\beta=2.3$
with local currents.}}
\begin{tabular}{@{}lclll}
\hline
\multicolumn{1}{c}{$\kappa$} & operator & $t_{min}=3$ & $t_{min}=4$ &
$t_{\rm min}=5$ \\
\hline
0.1925 & $\chi^{(sp)}$      & 0.183(14) & 0.207(27) & 0.19(5)     \\
0.1925 & $\chi^{(sp)}$ (*)  & 0.176(14) & 0.184(28) & 0.21(5)     \\
0.1925 & $\chi^{(sp)}$ (**) & 0.146(11) & 0.159(25) & 0.139(45)     \\
0.1925 & $T^{(loc)}_0$             &           & 0.29(6)   & 0.22(6)     \\ 
\hline
0.194  & $\chi^{(sp)}$      & 0.202(15) & 0.176(33) & 0.186(6)     \\  
0.194  & $T^{(loc)}_0$             &           & 0.27(9)   & 0.30(11)     \\ 
\hline
0.1955 & $\chi^{(sp)}$      & 0.179(10) & 0.170(21) & 0.170(45)     \\
\hline
\end{tabular} \\
\parbox{11cm}{* \footnotesize With plaquette field tensor.}\\
\parbox{11cm}{** \footnotesize With plaquette field tensor and smearing parameters B.}\\
\end{center}
\end{table*}

\newpage

\appendix\begin{center}  {\Large\bf Appendix}  \end{center}

\section{Discrete symmetries}

The following Hermitian representation of the Euclidean $\gamma$-matrices
is adopted:
\be
\gamma_0=\left(
\begin{array}{cc}
0 & 1\!\!1 \\
1\!\!1 & 0
\end{array} \right),
\ \ \ \
\gamma_k= -i\left(
\begin{array}{cc}
0 & \sigma_k \\
-\sigma_k & 0
\end{array}
 \right)\ ,
\ee
with anti-commutation property:
\be
\{\gamma_\mu,\gamma_\nu\}=2\,\delta_{\mu\nu}\ .
\ee
The matrix $\gamma_5$ is defined as
\be
\gamma_5=\gamma_1\gamma_2\gamma_3\gamma_0=\left(
\begin{array}{cc}
1\!\!1 & 0 \\
0 & -1\!\!1 
\end{array}
\right)\ ,
\ee
and the anti-Hermitian matrix $\sigma_{\mu\nu}$ reads
\be\label{sigma}
\sigma_{\mu\nu}=\frac{1}{2}[\gamma_{\mu},\gamma{_\nu}]\ .
\ee

\subsection{Parity \ppp} 

Denoting with 
$x_\spp$ the transformed 
coordinates, $x_\spp=(x_0,-x_1,-x_2,-x_3)$, 
the field transformations are:

\begin{eqnarray}
\lambda^{\spp}(x)&=&\gamma_0\lambda(x_\spp)\\
\blambda^{\spp}(x)&=&{\blambda}(x_\spp)\gamma_0\\
U_0^{\spp}(x)&=&U_0(x_\spp)\\
U_i^{\spp}(x)&=&U_i^\dagger(x_\spp-\hat{i})\ .
\end{eqnarray}

\subsection{Time reversal  \ttt}

In this case 
$x_\stt=(-x_0,x_1,x_2,x_3)$, and the field transformations are:
\begin{eqnarray}
\lambda^{\stt}(x)&=&\gamma_0\gamma_5\lambda(x_\stt)\\
\blambda^{\stt}(x)&=&{\blambda}(x_\stt)\gamma_5\gamma_0\\
U_0^{\stt}(x)&=&U_0^\dagger(x_\stt-\hat{0})\\
U_i^{\stt}(x)&=&U_i(x_\stt)\ . \\
\end{eqnarray}

The clover-symmetrized field tensor $P^{(cl)}_{\mu\nu}(x)$ of 
Eq.~(\ref{clover}) has the same transformation properties under \ppp\
and \ttt\ as its continuum counterpart $F_{\mu\nu}(x)$.
This does not apply to the simple plaquette lattice field tensor 
$P^{(pl)}_{\mu\nu}(x)$ of Eq.~(\ref{plaquette}).

\subsection{Charge conjugation  \ccc}

The invariance of the gluino field under charge conjugation \ccc\ 
(Majorana condition) reads
\begin{eqnarray}
\nonumber
\lambda(x)&=&\lambda^{\scc}(x)\:=\:C\blambda^{T}(x)\ , \\
\blambda(x)&=&\blambda^{\scc}(x)\:=\:\lambda^{T}(x)C^{-1}\ . 
\label{CC}\end{eqnarray}
The spinorial matrix $C$ is defined as
\be
C\equiv\gamma_0\gamma_2=\left(
\begin{array}{cc}
i\sigma_2 & 0 \\
0 & -i \sigma_2
\end{array} \right)\ ,
\ee
with the properties
\be
C^{-1}=-C=C^{T}\ 
\ee
and 
\be
C^{-1}\gamma_{\mu}C = -\gamma_{\mu}^T\; ,\;\;\;\;
C^{-1}\sigma_{\mu\nu}C=-\sigma_{\mu\nu}^T\ ,\;\;\;\; 
C^{-1}\gamma^5 C=\gamma^5\; .
\ee

\section{Renormalization of $O_{11/2}(x)$}

We discuss here the renormalization of the composite operator
$O_{11/2}(x)$ defining the continuum limit of $X(x)$ 
according to power-counting \cite{CuVe}, see Eq.~(\ref{renorm}).
In this context the particular lattice form of the operators is
immaterial and continuum notation will be used for simplicity.

In the present work we consider gauge invariant correlation functions
of the operator $O_{11/2}(x)$. In this case one can restrict the analysis
to mixing with gauge invariant operators. These operators must have in addition the
same transformation properties as $O_{11/2}(x)$ under the hypercubic 
group and the discrete symmetries \ccc, \ppp\ and \ttt.
Finally, dimensional considerations restrict the search to 
operators of dimension $d<11/2$ (dimension-$11/2$ or higher mixing 
can be neglected as clarified in 
Sec.~\ref{sec:renorm}).

The operator $O_{11/2}(x)$ transforms under $O(4)$ like the gluino field
\be 
\label{hctrans}
\lambda_\alpha(x)\rightarrow\sum_\beta S_{\alpha\beta}(R)\lambda_\beta(x)\ \ \ \
S(R) = \exp\{-\frac{1}{2} \sum_{\mu<\nu}\omega_{\mu\nu}\sigma_{\mu\nu}\}\ .
\ee
The connection with $O(4)$ is given by
\be
S(R)^{-1}\gamma_\mu S(R)=\sum_{\nu}R_{\mu\nu}\gamma_\nu\ .
\ee
The coefficient $\omega_{\mu\nu}$ represents the rotation angle in the plane $(\mu,\nu)$;
the hypercubic group, a discrete subgroup of O(4), is obtained by restricting
these angles to multiples of $\pi/2$.

Gauge invariant operators are obtained by taking the trace in color space
of products of fields in the adjoint representation. 
Available operators of this kind are the gluino field itself, the field tensor
$F_{\mu\nu}(x)$ and the covariant derivative in the adjoint representation
$D_\mu$. Dirac and tensorial indices should be combined in such a way that
the resulting operator transforms like~(\ref{hctrans}) 
under hypercubic transformations.
By using the algebra of the $\Gamma$-matrices
\be\label{gamma}
\Gamma=\{\one,\gamma^5,\gamma_{\mu},\gamma^5\gamma_{\mu},\sigma_{\mu\nu}\}
\ee
pairs of gluino fields can be assembled into bilinear expressions of the form
\be
\label{block}
O^{bilin}_{...}(x)\:=\:\blambda(x)\Gamma_{...}\,\lambda(x)
\ee
where Dirac indices are contracted and
the dots indicate possible tensorial indices of the $\Gamma$-matrix.
The transformation properties of the bilinears~(\ref{block}) 
depend on the $\Gamma$-matrix.
They belong to irreducible representations of both $O(4)$ and the
hypercubic group, see e.g.~\cite{MaZwGo83},  
\begin{itemize}
\item[] 
$\Gamma=\one,\gamma^5$: scalar, $I^{(+,-)}$ ($\tau_{1,4}^{(1)}$)\ ,
\item[] 
$\Gamma=\gamma_{\mu},\gamma^5\gamma_{\mu}$: vectorial, $(1/2,1/2)^{(+,-)}$ 
($\tau_{1,4}^{(4)}$)\ ,
\item[]
$\Gamma=\sigma_{\mu\nu}$: tensorial, $(1,0)\oplus(0,1)$ ($\tau_{1}^{(6)}$)\ .
\end{itemize}
The representations of $O(4)$ are indicated in the notation of~\cite{MaZwGo83}. 
In parenthesis we report the corresponding representations of the 
hypercubic group in the notation used in~\cite{GoetAl96} (see the following).

\vspace{.2cm}

{\bf General classification.} The general operator can be classified 
according to the number of contained gluino fields $\lambda(x)$, field tensors $F_{\mu\nu}(x)$ 
and covariant derivatives $D_\mu$, $n_{\lambda}$, $n_{F}$ and $n_{D}$.
In the case of an even number of gluino fields, one can exploit the algebra of 
the $\Gamma$-matrices and build bilinears of the form~(\ref{block}) with
no unpaired gluino field. Consequently all Dirac indices are contracted.
Products made up by these bilinears $O^{bilin}_{...}(x)$,  
field tensors $F_{\mu\nu}(x)$ and covariant derivatives $D_\mu$
transform according to a general tensorial representation of the hypercubic group
$O_{\mu_1,\ldots,\mu_n}$. An example is given by the dimension-4 operator
\be
O_{\mu_1\mu_2}=
\tr\left[D_{\mu_1}(\blambda(x)\gamma_{\mu_2}\lambda(x))\right] \ .
\ee

The decomposition into irreducible representations of the hypercubic 
group of tensorial representations has been studied in~\cite{GoetAl96}.
Of course the representation~(\ref{hctrans}) cannot be contained in these
representations.\footnote{This is evident for example if
one considers that a rotation of 2$\pi$ according to~(\ref{hctrans})
is not the identical transformation as in the tensorial representations.
Relation~(\ref{hctrans}) defines a double-valued irreducible
representation of $O(4)$ and of the hypercubic group.}
One can consequently restrict the investigation to operators containing
an odd number of gluino fields.
An additional restriction to the possible structures of mixing operators
comes from the general relation 
\be\label{relation}
\tr\left[D_{\mu_1}\cdots D_{\mu_n}A(x)\right]\:=\:
\partial_{\mu_1}\cdots\partial_{\mu_n}\tr\left[A(x)\right]\ ,
\ee
holding for a generic adjoint field $A(x)$.
Since $\tr[\lambda(x)]=0$ relation~(\ref{relation}) excludes cases 
where $n_{\lambda}=1$ and $n_{F}=0$.
This leaves only three possibilities for $d\leq 9/2$:
\begin{itemize}
\item[a)] 
$n_{\lambda}=3$, $n_{F}=0$ , $n_{D}=0$ \ \ \ \     ($d=9/2$)\ ,
\item[b)]
$n_{\lambda}=1$, $n_{F}=1$ , $n_{D}=1$ \ \ \ \    ($d=9/2$)\ ,
\item[c)] 
$n_{\lambda}=1$, $n_{F}=1$ , $n_{D}=0$ \ \ \ \    ($d=7/2$)\ .
\end{itemize}

\subsection{$n_{\lambda}=3$, $n_{F}=0$ , $n_{D}=0$}

The most general gauge invariant three-gluino operator can be 
expressed as
\be\label{tensorial}
O_{\alpha, ...}(x)=
\tr\left[(\blambda(x)\Gamma_{...}\lambda(x))\,(\Gamma^\prime_{...}\lambda(x))_\alpha\right]\ ,
\ee
where as usual the dots indicate possible tensorial indices.
Dirac indices are contracted in the bilinear.
The transformation rule in the free Dirac index $\alpha$
is the right one given by Eq.~(\ref{hctrans}). This Dirac index will be
suppressed in the future notation. 
$O_{\alpha, ...}(x)$ defines a tensorial representation of $O(4)$ 
(and of the hypercubic group) in the remaining tensorial indices.
Imposing invariance under $O(4)$ is equivalent to requiring that
these indices are contracted so as to obtain a scalar. 
The procedure is standard and the result is 
\begin{eqnarray}
\label{3glop_first}
O_{S}(x)&\!\!=\!\!&\tr\left[(\blambda(x)\lambda(x))\lambda(x)\right]  \\
O_{P}(x)&\!\!=\!\!&\tr\left[(\blambda(x)\gamma^5\lambda(x))\gamma^5\lambda(x)\right]  \\
\label{3glop_vec}
O_{V}(x)&\!\!=\!\!&\tr\left[\sum_{\mu}(\blambda(x)\gamma_\mu\lambda(x))\gamma_\mu\lambda(x)\right]  \\
O_{A}(x)&\!\!=\!\!&\tr\left[\sum_{\mu}(\blambda(x)\gamma^5\gamma_\mu\lambda(x))\gamma^5\gamma_\mu\lambda(x)\right]  \\
\label{3glop_last}
O_{T}(x)&\!\!=\!\!&\tr\left[\sum_{\mu\nu}(\blambda(x)\sigma_{\mu\nu}\lambda(x))\sigma_{\mu\nu}\lambda(x)\right] \ .
\end{eqnarray}
For Majorana fermions any three-gluino operator complying with $O(4)$ 
invariance, (\ref{3glop_first})-(\ref{3glop_last}), vanishes.
The Majorana condition~(\ref{CC})  can be directly used 
to show vanishing of $O_{S}(x)$, $O_{P}(x)$ 
and  $O_{A}(x)$:
\be
O_{S}(x)=O_{P}(x)=O_{A}(x)=0\ .
\ee
Fierz rearrangements
\begin{eqnarray}
O_{V}(x)&\!\!=\!\!&-O_{A}(x)\:=\:O_{S}(x)-O_{P}(x)\label{zero1}\ , \\
O_{T}(x)&\!\!=\!\!&-O_{S}(x)-O_{P}(x)\ \label{zero2}
\end{eqnarray}
imply vanishing of the remaining two operators $O_{V}(x)$
and $O_{T}(x)$.

The discussion is however not complete since the true symmetry of the lattice
is the hypercubic one. So the question arises whether tensorial indices 
can be combined in~(\ref{tensorial}) in a different way 
from~(\ref{3glop_vec})-(\ref{3glop_last}), 
while still complying with the hypercubic invariance. 
The resulting operator would represent a potential
Lorentz-breaking term in the renormalization of $O_{11/2}(x)$.
The argument can be made more rigorous by considering that
Eq.~(\ref{tensorial}) defines tensorial representations of the hypercubic group.
The question is whether singlet representations of the hypercubic group, 
which are not $O(4)$ scalars, are
contained in the tensorial representations~(\ref{tensorial}).
We rely in this on the detailed discussion of the subject contained in~\cite{GoetAl96}. 

Singlet representations of the hypercubic group which are not 
necessarily $O(4)$ scalars
are contained in tensorial representations with even number 
of indices $n\geq4$.
In the classification~(\ref{tensorial}) the only possible candidate is
the operator with $\Gamma_{\mu\nu}=\Gamma^\prime_{\mu\nu}=\sigma_{\mu\nu}$
containing four tensorial indices
\be\label{tensorial4}
(O^{(4)}_T)_{\mu_1\mu_2\mu_3\mu_4}(x)=\tr\left[(\blambda(x)\sigma_{\mu_1\mu_2}
\lambda(x))\sigma_{\mu_3\mu_4}\lambda(x)\right]\ .
\ee
This means that $O(4)$-breaking versions of the operator
$O_{T}(x)$, Eq.~(\ref{3glop_last}), are in principle possible. 
In order to decide the question we consider the general case 
of an operator with four tensorial indices 
$O^{(4)}_{\mu_1\mu_2\mu_3\mu_4}(x)$.
One-dimensional representations of the hypercubic group 
which do not coincide with O(4) scalars are given by~\cite{GoetAl96}
\begin{eqnarray}
&& \ \ \ \sum_{\mu} O^{(4)}_{\mu,\mu,\mu,\mu}(x)\ \ \ \ \ \ (\tau_1^{(1)})\ , \label{new1}\\
&& \ \ \ O^{(4)}_{\{0,1,2,3\}}(x)\ \ \ \ (\tau_2^{(1)})\ . \label{new2}
\end{eqnarray}
The symbol $\{...\}$ in (\ref{new2}) defines complete symmetrization 
on the Lorentz indices.
In the case we are interested in, ($O^{(4)}_T)_{\mu_1\mu_2\mu_3\mu_4}(x)$,
antisymmetry of $\sigma_{\mu\nu}$ implies trivially vanishing of these new combinations
of indices (combination (\ref{new2}) would have the wrong symmetry anyway).
Consequently no three-gluino operators complying with hypercubic invariance can be built.

\subsection{$n_{\lambda}=1$, $n_{F}=1$ , $n_{D}=1$}

Gauge invariance imposes the form
\be
O(x)_{\mu_1,\mu_2,\mu_3,...}=\tr\left[D_{\mu_1} \{F_{\mu_2\mu_3}(x)\Gamma_{...}\, \lambda(x)\right\}]\ .
\ee
The symbolic expression $D_\mu\{ABC \cdots\}$ 
is a collective representation of all the
ways the covariant derivative $D_\mu$ can act on the
products of the adjoint fields $A,B,C \ldots$. Since the adjoint 
covariant derivative satisfies the Leibniz rule,
it is sufficient to consider only operators where $D_\mu$ acts on just
one field $A,B,C \ldots$.
Again, tensorial indices must be combined as to obtain singlets
under the hypercubic group. $O(4)$-invariance and \ppp\ leave only two possibilities
\begin{eqnarray}\label{case1}
&&\tr\left[D_\mu \{F_{\mu\nu}(x)\gamma_{\nu}\lambda(x)\}\right]\ \ \ \ \ \ \ \ \ \ \ \ \ (i)
\\\label{case2}
&&\tr\left[\epsilon_{\mu\nu\rho\sigma}D_\mu \{F_{\nu\rho}(x)\gamma_5\gamma_{\sigma}\lambda(x)\}\right]\ \ \ \  (ii)\ .
\end{eqnarray}
Also in this case new hypercubic-invariant combinations could come
only from 4-index tensors, e.g.
\be
O(x)_{\mu_1,\mu_2,\mu_3,\mu_4}=\tr\left[D_{\mu_1} \{F_{\mu_2\mu_3}(x)\gamma_{\mu_4}\lambda(x)\right\}]\  .
\ee
An analysis analogous to the one performed in the previous
Subsection leads to the conclusion that no new, non-trivial operators
arise from combinations~(\ref{new1}),(\ref{new2}).
In this case antisymmetry of $F_{\mu\nu}(x)$ plays the key r\^ole.
So we concentrate on the Lorentz-conserving operators~(\ref{case1}) and~(\ref{case2}). 

Case (i). The first possibility according to the Leibniz rule
for the covariant derivative in~(\ref{case1}) 
\be\label{poss1}
\tr\left[(D_\mu F_{\mu\nu}(x))\gamma_{\nu}\lambda(x)\right]
\ee
vanishes on-shell. Indeed, using the equation of motion 
for $F_{\mu\nu}(x)$ it can be rewritten as a three-gluino operator, vanishing 
identically according to the previous discussion.
The second possibility for the Leibniz rule
\be
\tr\left[F_{\mu\nu}(x)\gamma_{\nu}D_\mu\lambda(x)\right] 
\ee
is equivalent on shell to
\be
\tr\left[D_\mu(F_{\mu\nu}(x)\gamma_{\nu}\lambda(x))\right]\ ,
\ee
given the vanishing of~(\ref{poss1}).
Using relation~(\ref{relation}) we arrive at the operator  
$\partial_\mu T_\mu(x)$.

Case (ii). This goes along the same lines as case (i).
The relation 
\be
\epsilon_{\mu\nu\rho\sigma}\gamma^5\gamma_\sigma=
-\sigma_{\nu\rho}\gamma_\mu+(\delta_{\mu\rho}\gamma_{\nu}-
\delta_{\mu\nu}\gamma_{\rho})
\ee
can be used. The part containing the $\gamma_\mu$ matrices 
reduces to the case (i) already considered. The new combination is:
\be
\tr\left[D_\mu \{F_{\nu\rho}(x)\sigma_{\nu\rho}\gamma_\mu\lambda(x)\}\right]\ .
\ee
Again, there are essentially two possibilities:
\be
\tr\left[F_{\mu\nu}(x)\sigma_{\mu\nu}D\!\!\!\!/\lambda(x)\right]
\ee
and
\be
\tr\left[D_\mu (F_{\nu\rho}(x)\sigma_{\nu\rho}\gamma_\mu\lambda(x))\right]\ .
\ee
On-shell, the first operator reduces to the lower dimensional operator 
$\chi(x)$~(\ref{chi}) due to the equation of motion for the field $\lambda(x)$,
and can be neglected in this discussion.
Rule~(\ref{relation}) shows that the second operator is just
the divergence of the SUSY current $\partial_\mu S_\mu(x)$.

In summary, the systematic scan of all possible dimension-$9/2$ operators with the
required symmetry properties and the on-shell restriction leaves us with the
two operators $\partial_\mu S_\mu(x)$ and $\partial_\mu T_\mu(x)$.

\subsection{$n_{\lambda}=1$, $n_{F}=1$ , $n_{D}=0$}

Using gauge invariance one gets of the general form
\be
O(x)_{\mu_1,\mu_2,...}=\tr\left[F_{\mu_1\mu_2}(x)\Gamma_{...}\lambda(x)\right]\ .
\ee
The only hypercubic-invariant combination of indices is
\be
\tr\left[F_{\mu\nu}(x)\sigma_{\mu\nu}\lambda(x)\right]=\chi(x)\ .
\ee
Again, no Lorentz-breaking combination appears.

This exhausts all possibilities for the power subtractions of $O_{11/2}(x)$
and implies the form~(\ref{renormward}) of the SUSY WIs. 

\section{Insertion operators}

In this Appendix we use for simplicity, as in the previous one, 
notions of the continuum. Even if the true symmetry of the lattice
is the hypercubic one, the analysis is carried over by
using the more restrictive Lorentz invariance. 
Indeed Lorentz breaking terms should not be considered, 
given that $O(a)$ effects are neglected.

Consider an insertion operator transforming under a generic representation 
of the Lorentz group (we restrict the discussion 
to operators depending on one coordinate).
Due to the spinoral character of the SUSY WIs we need spinorial operators.
Operators of this type are
\be\label{genins}
{\cal O}_{\mu_1\ldots\mu_n,\alpha}(x)=T_{\mu_1\ldots\mu_n}(x)\,\psi_{\alpha}(x)\ ,
\ee
where $T_{\mu_1\ldots\mu_n}(x)$ is 
an operator transforming according to the tensorial representation
of the Lorentz group and $\psi_{\alpha}(x)$ is a bispinor.
Irreducible representations are projected-out by
suitable (anti)symmetrizations, extraction of traces on the free indices.
We consider zero spatial momentum WIs~(\ref{zeromomward}). These are obtained by 
taking the vacuum expectation value of operators like e.g.
\be\label{composition}
{\cal O}_{\mu_1\ldots\mu_n,\alpha\beta}(x_0,y_0)=
\frac{1}{V_s}\int d\vec{x}d\vec{y}\:(\partial_0 S_0)_\alpha(\vec{x},x_0)\, 
{\cal O}_{\mu_1\ldots\mu_n,\beta}(\vec{y},y_0)\ .
\ee
The total angular momentum $J$ of the above operator results from 
the composition of the spins of the two operators 
$\int d\vec{x}\, (\partial_0 S_0)_\alpha(\vec{x},x_0)$ and
$\int d\vec{x}\, {\cal O}_{\mu_1\ldots\mu_n\beta}(\vec{x},y_0)$; the `orbital'
angular momentum is zero because of the summation on space coordinates.
In order to get non-trivial WIs one must have of course $J=0$.  
Since the operator $\partial_\mu S_\mu(x)$ is a bispinor,
the condition $J=0$ implies that the insertion operator must
contain at least one spin-$1/2$ component.
The irreducible representations of the Lorentz group\footnote{As usual we 
take into account also \ppp.} with this property are of the type
$((S,S^{\prime})\oplus (S^{\prime},S))$, with $|S-S^{\prime}|=1/2$.

The lowest dimensional gauge invariant operator 
has dimension $d=7/2$. It has the form~(\ref{genins}) with
tensorial part given by $F_{\mu\nu}(x)$ and $\psi(x)=\lambda(x)$. 
Its transformation properties under the Lorentz group are given by
\be\label{decomp}
((1,0)\oplus (0,1))\otimes((\frac{1}{2},0)\oplus (0,\frac{1}{2}))=
((\frac{1}{2},0)\oplus (0,\frac{1}{2}))\oplus (\frac{1}{2},\frac{1}{2})
\oplus (\frac{1}{2},\frac{1}{2})\oplus ((\frac{1}{2},1)\oplus (1,\frac{1}{2})) \ .
\ee
We see that two (and only two) representations on the r.h.s. of the above decomposition 
contain spin-$1/2$, namely
$((\frac{1}{2},0)\oplus (0,\frac{1}{2}))$ and $((\frac{1}{2},1)\oplus (1,\frac{1}{2}))$.
The first component is given by the operator $\chi(x)$, the second is 
present for example in $S_\mu(x)$ and $T_\mu(x)$
transforming like
\be
(\frac{1}{2},\frac{1}{2})\otimes((\frac{1}{2},0)\oplus (0,\frac{1}{2})) = 
((\frac{1}{2},0)\oplus (0,\frac{1}{2}))\oplus
((\frac{1}{2},1)\oplus (1,\frac{1}{2}))\ .
\ee 
Pure spin-$1/2$ operators are given by $\chi(x)$, 
$S_0(x)$ and $T_0(x)$, while
$S_i(x)$ and $T_i(x)$ contain also spin-$3/2$ components.
Eq.~(\ref{decomp}) implies that any spin-$1/2$ component of any dimension-$7/2$ operator
can be expressed as a linear combination of two operators chosen as a basis, for example 
$\chi(x)$ and $T_0(x)$ (see e.g. relations~(\ref{oprel1}), (\ref{oprel2})).
Consequently only two operators give independent WIs 
at the lowest dimension.

For a given insertion operator the WIs consist of a number of independent equations
which equals the number of rotational invariant (spin-0) 
components contained in the 
operatorial product~(\ref{composition}).
A straightforward analysis including also \ppp\
reveals that these are two for ${\cal O}\equiv\chi(x), S_0(x), T_0(x)$,
as given in~(\ref{comp1}) and (\ref{comp2}).
It should be stressed that considering other insertion operators
like for example $S_i(x)$ or $T_i(x)$ would not give additional information
to that given by e.g. $\chi(x)$, $T_0(x)$.
One would get only different combinations of the same on-shell WIs. 

The discrete symmetries \ttt\ and \ccc\  imply
relations for the correlations involved in the SUSY WIs.
For the correlations defined in~(\ref{comp1}), (\ref{comp2}) these
are 
\begin{eqnarray}
\ttt: \quad  C^{(S,{\cal O})}(t)&=&\gamma_0\gamma_5C^{(S,{\cal O})}(-t)\gamma_5\gamma_0\label{rel_t}\\
\ccc: \quad  C^{(S,{\cal O})}(t)&=&
\gamma_0\gamma_2\gamma_5(C^{(S,{\cal O})})^*(t)\gamma_5\gamma_2\gamma_0 \label{rel_c}\ .
\end{eqnarray}
In terms of the components defined in~(\ref{comp1}), 
relation~(\ref{rel_t}) reads
\begin{eqnarray}
C^{(S,{\cal O})}_{\one}(-t)&\!\!\!=\!\!\!&C^{(S,{\cal O})}_{\one}(t)\, \nonumber\\ 
C^{(S,{\cal O})}_{\gamma_0}(-t)&\!\!\!=\!\!\!&-C^{(S,{\cal O})}_{\gamma_0}(t)\ .
\label{reflection}
\end{eqnarray}
In the case  ${\cal O} \equiv S_0,T_0$ an extra minus sign must be included.
Relation~(\ref{rel_c}) ensures reality of the two components
\begin{eqnarray}
C^{(S,{\cal O})}_{\one}(t)&=&(C^{(S,{\cal O})}_{\one}(t))^*\nonumber\\
C^{(S,{\cal O}}_{\gamma_0}(t)&=&(C^{(S,{\cal O})}_{\gamma_0}(t))^*\ .
\label{reality}
\end{eqnarray}

Properties related to \ppp\ and \ttt\
apply unchanged for the lattice theory if the clover-symmetrized
lattice field tensor $P^{(cl)}_{\mu\nu}(x)$ is 
used for the insertion operators. 
The simple discretization $P^{(pl)}_{\mu\nu}(x)$ 
breaks \ppp\ and \ttt\ and additional 
$O(a)$ pseudoscalar components are present in the correlations.
In this case, for operators extended in the time-direction, the
time-reflection properties~(\ref{reflection}) are violated.

\section{Linear fit}

We perform a linear fit to solve the WI~(\ref{system}) 
for $Z_T Z_S^{-1}$ and $am_S Z_S^{-1}$
including several consecutive time-separations $(t_{min},\cdots, L_t/2$).
We define
\be
A=Z_T Z_S^{-1}, \ \ B=am_S Z_S^{-1}\ , 
\ee
and $x_{1,t}$, $y_{1,t}$ and $z_{1,t}$ as the different components of the correlation 
functions at different times (see Eqs.~(\ref{comp1}),(\ref{comp2}))
\be
\begin{array}{lll}
x_{1,t} = C^{(S,{\cal O})}_{\one}(t)\ , & y_{1,t} = C^{(T,{\cal O})}_{\one}(t)\ , & 
z_{1,t} = C^{(\chi,{\cal O})}_{\one}(t)\ , \\
x_{2,t} = C^{(S,{\cal O})}_{\gamma_0}(t)\ , & y_{2,t} = C^{(T,{\cal O})}_{\gamma_0}(t)\ , & 
z_{2,t} = C^{(\chi,{\cal O})}_{\gamma_0}(t)\ . \\
\end{array}
\ee
The overdetermined system reads
\be
\label{FITAPP}
 x_{i,t} +A y_{i,t} =Bz_{i,t}, \ \ \ \ i=1,2\ , \ \ \ t=t_{min},\cdots, L_t/2\ .
\ee
We get the best estimates for $Z_T Z_S^{-1}$ and $am_S Z_S^{-1}$ ($A$ and $B$)
by minimizing the quantity 
\be
H=\sum_{i=1}^2\sum_{t=t_{min}}^{L_t/2} (x_{i,t}+Ay_{i,t}-Bz_{i,t})^2\ .
\ee
With the conditions
\begin{eqnarray}
\frac{\partial H}{\partial A}=\hphantom{-} 2 \sum_{i,t} y_{i,t}\, (x_{i,t}+Ay_{i,t}-Bz_{i,t})=0 \\
\frac{\partial H}{\partial B}=-2 \sum_{i,t} z_{i,t}\, (x_{i,t}+Ay_{i,t}-Bz_{i,t})=0
\end{eqnarray}
and the definition
\be
\sum_{i,t}\, x_{i,t}\, y_{i,t} = <x,y>\ ,
\ee
$A$ and $B$ are given by
\begin{eqnarray}
A=\frac{<y,z><x,z>-<x,y><z,z>}{<y,y><z,z>-<y,z>^2}\ , \\
B=\frac{<x,z><y,y>-<x,y><y,z>}{<y,y><z,z>-<y,z>^2}\ .
\end{eqnarray}

\end{document}